%% file: SISR_TCSVT.tex
\documentclass[journal]{IEEEtran}
\usepackage[breaklinks=true,colorlinks, bookmarks=false]{hyperref}
\usepackage{color}
\usepackage{amsmath}
\usepackage{graphicx} 
\usepackage{multirow}
\usepackage{amssymb}
\usepackage{makecell}
\usepackage{booktabs}
\usepackage{bm}

\newcommand{\zjq}[1]{\textcolor[rgb]{0.0,0.0,0.0}{#1}}

\ifCLASSINFOpdf
\else
\fi
\begin{document}
\title{A Two-Stage Attentive Network for Single Image Super-Resolution}
\author{Jiqing Zhang, Chengjiang Long$^*$, Yuxin Wang$^*$, Haiyin Piao, Haiyang Mei, Xin Yang$^*$, Baocai Yin
	\thanks{This work was supported in part by the National Natural Science Foundation of China under Grant 91748104, Grant 61972067, Grant 61632006, in part by the National Key Research and Development Program of China under Grant 2018AAA0102003, and the Innovation Technology Funding of Dalian (Project No. 2018J11CY010, 2020JJ26GX036).}
	\thanks{Jiqing Zhang, Yuxin Wang, Haiyang Mei, Xin Yang, and Baocai Yin are with the Department of Electronic Information and Electrical Engineering,
		Dalian University of Technology, Dalian, 116024, China. 
		E-mail: \href{mailto:jqz@mail.dlut.edu.cn}{jqz@mail.dlut.edu.cn}; \href{mailto:wyx@dlut.edu.cn}{wyx@dlut.edu.cn}; \href{mailto:mhy666@mail.dlut.edu.cn}{mhy666@mail.dlut.edu.cn}; \href{mailto:xinyang@dlut.edu.cn}{xinyang@dlut.edu.cn}; \href{mailto:ybc@dlut.edu.cn}{ybc@dlut.edu.cn}.}% <-this % stops a space
	\thanks{Chengjiang Long is with JD Finance America Corporation, CA, USA 94043. E-mail: \href{mailto:cjfykx@gmail.com}{chengjiang.long@jd.com}.}% <-this % stops a space
	\thanks{Haiyin Piao is with  School of Electronics and Information, Northwestern Polytechnical University, Xian, China. E-mail: \href{mailto:haiyinpiao@mail.nwpu.edu.cn}{haiyinpiao@mail.nwpu.edu.cn}.}
	% \thanks{$^*$Xin Yang (\href{mailto:xinyang@dlut.edu.cn}{xinyang@dlut.edu.cn}), Chengjiang Long (\href{mailto:chengjiang.long@jd.com}{chengjiang.long@jd.com}), and Yuxin Wang (\href{mailto:wyx@dlut.edu.cn}{wyx@dlut.edu.cn}) are the corresponding authors.}
	\thanks{$^*$Xin Yang (\href{mailto:xinyang@dlut.edu.cn}{xinyang@dlut.edu.cn}), Chengjiang Long, and Yuxin Wang are the corresponding authors.}
	\thanks{Copyright © 2021 IEEE. Personal use of this material is permitted. However, permission to use this material for any other purposes must be obtained from the IEEE by sending an email to pubs-permissions@ieee.org.}
}
%\author{Michael~Shell,~\IEEEmembership{Member,~IEEE,}
%        John~Doe,~\IEEEmembership{Fellow,~OSA,}
%        and~Jane~Doe,~\IEEEmembership{Life~Fellow,~IEEE}% <-this % stops a space


\markboth{IEEE Transactions on Circuits and Systems for Video Technology}%
{Shell \MakeLowercase{\textit{et al.}}: Bare Demo of IEEEtran.cls for IEEE Journals}



\maketitle

\begin{abstract}
Recently, deep convolutional neural networks (CNNs) have been widely explored in single image super-resolution (SISR) and contribute remarkable progress. However, most of the existing
CNNs-based SISR methods do not adequately explore contextual information in the feature extraction stage and pay little attention to the final high-resolution (HR) image reconstruction step, hence hindering the desired SR performance. To address the above two issues, in this paper, we propose
a two-stage attentive network (TSAN) for accurate SISR in a coarse-to-fine manner. Specifically, we design a novel multi-context attentive block (MCAB) to make the network focus on more informative contextual features. Moreover, we present an essential refined attention block (RAB) which could explore useful cues in HR space for reconstructing fine-detailed HR image. Extensive evaluations on four benchmark datasets demonstrate the efficacy of our proposed TSAN in terms of quantitative metrics and visual effects. Code is available at \url{https://github.com/Jee-King/TSAN}.
\end{abstract}


\input{introduction}
\input{related_works}

\input{methodology}
\input{experiments}
\input{conclusion}

% \fi
\IEEEpeerreviewmaketitle

\ifCLASSOPTIONcaptionsoff
  \newpage
\fi

\clearpage

\begin{IEEEbiography}[{\includegraphics[width=1in,height=1.25in,clip,keepaspectratio]{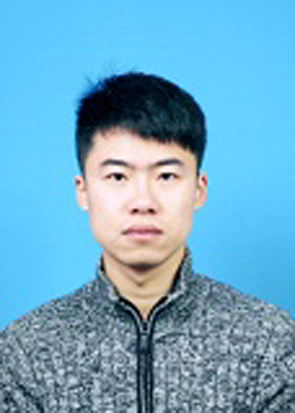}}]{Jiqing Zhang}
	received the B.Eng. degree in computer science and technology from Dalian Maritime University, Dalian, China, in 2017. 
	He is currently	pursuing the Ph.D. degree in computer application technology at Dalian University of Technology, Dalian, China.
	His research interests include computer vision, machine learning, and image processing.
\end{IEEEbiography}

\begin{IEEEbiography}[{\includegraphics[width=1in,height=1.25in,clip,keepaspectratio]{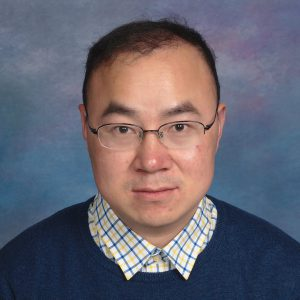}}]{Chengjiang Long}
	is currently a Principal Scientist in JD Finance America Corporation (a part of JD.COM) since June 2020. Prior to working at JD.COM, he worked as a Computer Vision Researcher/Senior R\&D Engineer at Kitware from February 2016 to April 2020. He also worked as an Adjunct Professor at University at Albany, SUNY from August 2018 to May 2020, and was an Adjunct Professor at Rensselaer Polytechnic Institute (RPI) from Jan 2018 to May 2018. He received the M.S. degree in Computer Science from Wuhan University in 2011 and a B.S degree in Computer Science and Technology from Wuhan University in 2009. He got his Ph.D. degree in Computer Science from Stevens Institute of Technology in 2015. During his Ph.D. study, he worked at NEC Labs America and GE Global Research as a research intern in 2013 and 2015, respectively. To date, he has published 50 papers including top journals such T-PAMI, IJCV, T-IP and T-MM, top international conferences such as CVPR, ICCV, and AAAI, and owns 1 patent. He is also the reviewer for more than 20 top international journals and conferences. His research interests involve various areas of computer vision, computer graphics, multimedia, machine learning, and artificial intelligence. He is a member of IEEE and AAAI.
\end{IEEEbiography}

\begin{IEEEbiography}[{\includegraphics[width=1in,height=1.25in,clip,keepaspectratio]{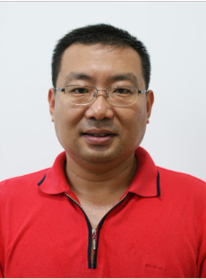}}]{Yuxin Wang}
	is an associate professor 
	and master tutor of Dalian University of Technology. His main research interests include parallel and distributed computing, big data analysis and application.
\end{IEEEbiography}

\begin{IEEEbiography}[{\includegraphics[width=1in,height=1.25in,clip,keepaspectratio]{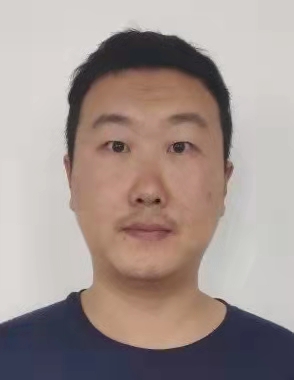}}]{Haiyin Piao}
	is currently working toward the Dr.Eng. degree in electronics and information at Northwestern Polytechnical University (NWPU), Xian, China.	
	He is currently also a Senior Engineer at Artificial Intelligence Laboratory, SADRI Institute, Shenyang, China. His research interests include multi-agent reinforcement learning, game theory with particular attention to aerospace applications.
\end{IEEEbiography}

\begin{IEEEbiography}[{\includegraphics[width=1in,height=1.25in,clip,keepaspectratio]{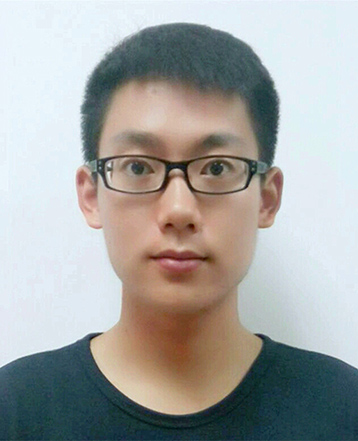}}]{Haiyang Mei}
	is a PH.D. student in the Department of Computer Science at Dalian University of
	Technology, China. His research interests include computer vision, machine learning, and image processing.
\end{IEEEbiography}

\begin{IEEEbiography}[{\includegraphics[width=1in,height=1.25in,clip,keepaspectratio]{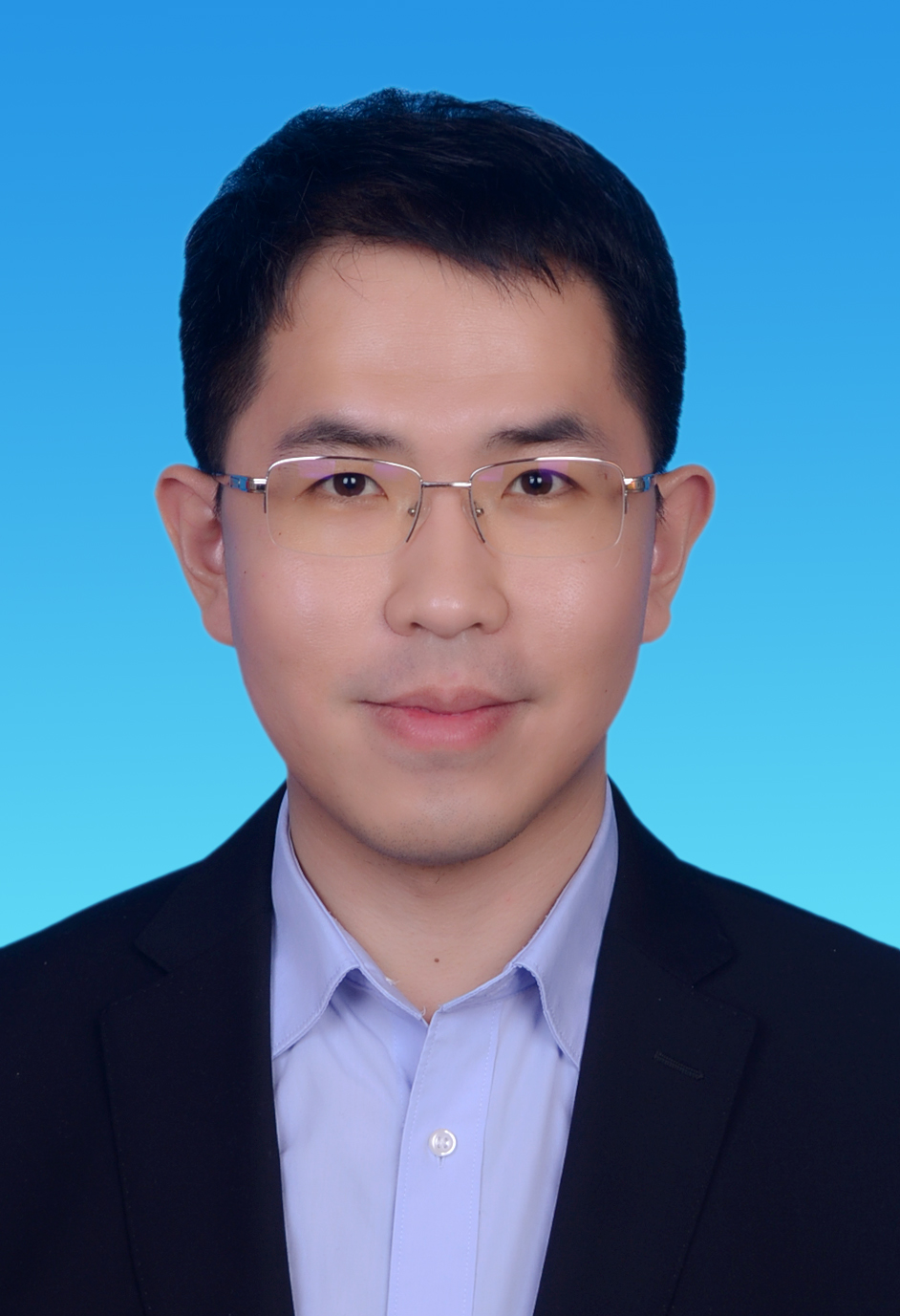}}]{Xin Yang}
	is a Professor in the Department of Computer Science at Dalian University of Technology, China. Yang received his B.S. degree in Computer Science from Jilin University in 2007.
	From 2007 to June 2012, he was a joint Ph.D. student at Zhejiang University and UC Davis for Graphics and received his Ph.D. degree in July 2012. His research interests include computer graphics and robotic vision.
\end{IEEEbiography}

\begin{IEEEbiography}[{\includegraphics[width=1in,height=1.25in,clip,keepaspectratio]{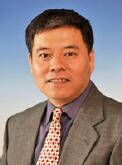}}]{Baocai Yin}
	is a Professor of Computer Science at Dalian University of Technology and the Dean of
	the Faculty of Electronic Information and Electrical Engineering. His research concentrates on digital	multimedia and computer vision. He received his B.S. degree and Ph.D. degree in Computer Science, each from Dalian University of Technology.
\end{IEEEbiography}

%\end{IEEEbiography}
%% insert where needed to balance the two columns on the last page with
%%\newpage


%\vfill


\end{document}

%% file: introduction.tex
\section{Introduction}
 
\zjq{\IEEEPARstart{S}{ingle} Image Super-Resolution (SISR) refers to reconstructing a visually pleasing high-resolution (HR) image from a low-resolution (LR) one.} It is a fundamental topic in the computer vision community and is an intense demand for diverse applications such as medical imaging, security, and surveillance imaging. The key to SISR problem lies in how to effectively
extract useful information from the input image and how to leverage extracted features to reconstruct the fine-detailed HR image. Since multiple HR images can be downsampled to the same LR image and it is a one-to-many mapping relation to recover HR images from one LR image, SISR is an ill-posed and still challenging problem in spite that numerous methods have been proposed.

\def\wvisualimagex0{0.5\linewidth}
\def\hvisualimagex0{3.05in}
\def\wvisualx0{0.22\linewidth}
\def\hvisualx0{0.45in}
\begin{figure}[t]
	\setlength{\tabcolsep}{1.0pt}
	\small
	\centering
	\begin{tabular}{ccc}
		\multirow{12}{*}[25.5pt]{\includegraphics[width=\wvisualimagex0, height=\hvisualimagex0]{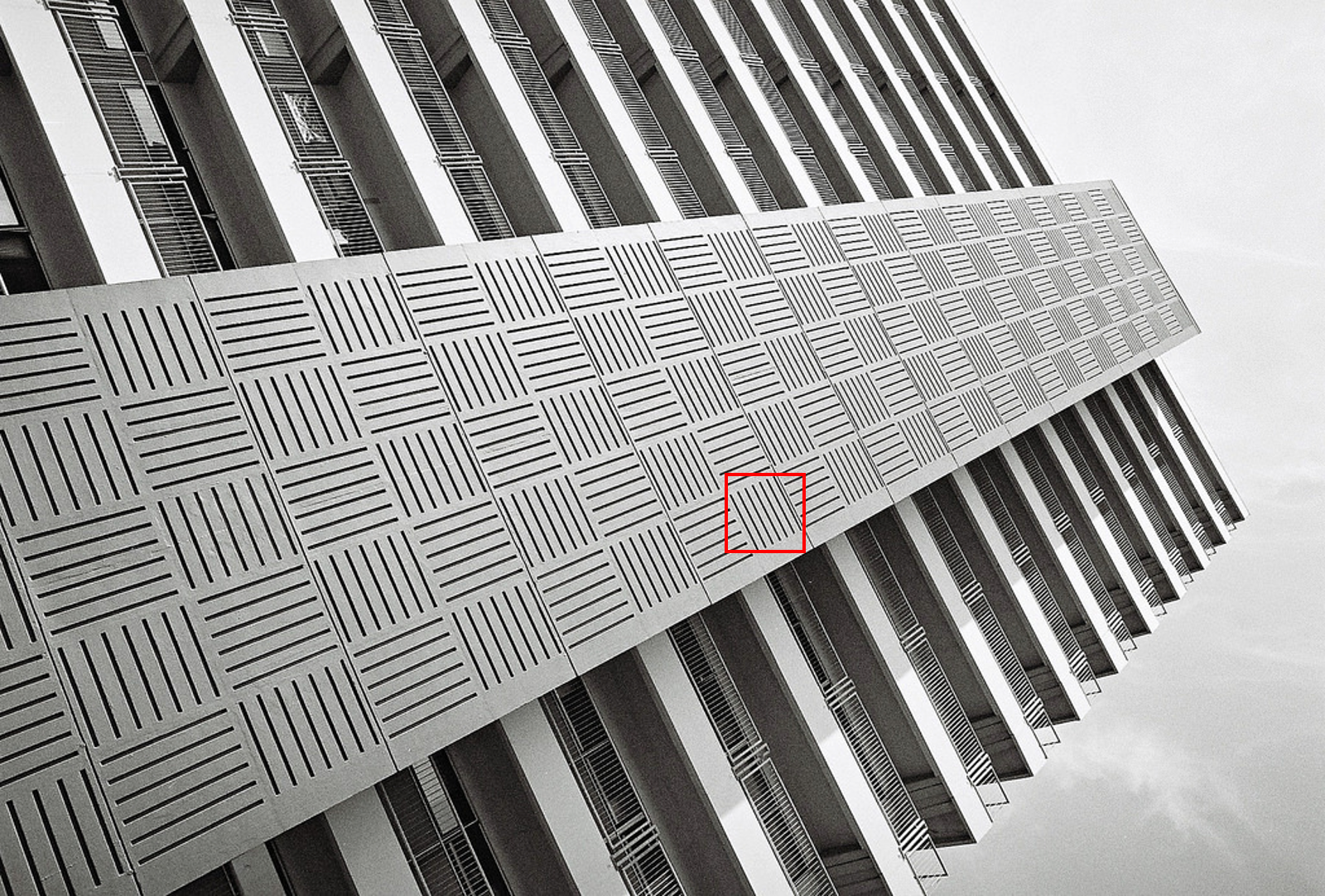}} & \includegraphics[width=\wvisualx0, height=\hvisualx0]{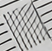} &
		\includegraphics[width=\wvisualx0, height=\hvisualx0]{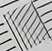} \\
		& EDSR~\cite{lim2017enhanced} & RDN~\cite{zhang2018residual} \\
		& (21.09/0.7591) &(20.88/0.7543) \\
		
		& \includegraphics[width=\wvisualx0, height=\hvisualx0]{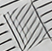} 
		& \includegraphics[width=\wvisualx0, height=\hvisualx0]{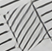} \\
		& FRSR~\cite{Soh_2019_CVPR}  & OISR~\cite{he2019ode} \\
		& (20.53/0.7432) & (20.66/0.7485)\\
		
		& \includegraphics[width=\wvisualx0, height=\hvisualx0]{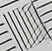}
		& \includegraphics[width=\wvisualx0, height=\hvisualx0]{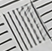} \\
		& RNAN~\cite{zhang2019rnan} & RCAN~\cite{zhang2018image} \\
		& (21.13/0.7600)  & (21.32/0.7708) \\

		& \includegraphics[width=\wvisualx0, height=\hvisualx0]{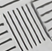}  
		& \includegraphics[width=\wvisualx0, height=\hvisualx0]{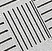} \\
		& Ours  & GT \\
	    & \textbf{(21.34/0.7711)} &(PSNR/SSIM) \\	
	\end{tabular}
	
	\caption{Visual comparison between different algorithms on \emph{img092} from  Urban100~\cite{huang2015single} with scale factor $\times$3. Our TSAN obtains better visual quality and recovers more image	details compared with other state-of-the-art SR methods.}
	\label{fig:teaser}
\end{figure}

\zjq{As a cutting-edge technique, deep-learning especially convolutional neural networks (CNNs) have been widely used to handle SISR~\cite{dong2016image,li2019srfbn,cai2019toward,yang2018drfn,he2019mrfn,yan2019deep,ning2020accurate,song2020efficient,songstereoscopic,zhang2020deep,dong2018denoising,yoo2020rethinking,soh2020meta,hu2019channel, chang2018energy,wang2018resolution,chang2018energy,li2018image,dong2018robust}.} In spite of remarkable progress achieved in SISR, existing CNN-based methods still face with three main limitations: 
(1) for feature extraction, early methods first apply interpolation strategies (\textit{e.g.}, bicubic) to process the input image to the desired size and then use CNNs to extract features from the upsampled image. As the interpolation often results in visible reconstruction artifacts, some models extract raw features directly from the input LR image and struggle to enhance the ability of feature extraction by simply deepening/widening the network. These methods blindly increase the depth of the network to enhance the performance of the network but ignore taking full use of the contextual information. As the depth of the network increases, the features gradually disappear in the process of transmission; 
(2) all features are treated equally in these methods, which weakens the discrimination ability of the network to extract more informative features. Although approaches~\cite{zhang2018image,dai2019second} retain some detailed information with channel attention, they struggle in preserving informative textures and restoring natural details since they 
ignore to explore the cross-dimension interaction; 
(3) for HR image reconstruction process, most models reconstruct HR image in one upsampling step at the end of the network, using features learned only in LR space. This setting would increase the difficulties of training for large scaling factors and make the network failed to explore useful cues in HR space for reconstructing visually pleasant HR image.

To address the above limitations, we propose a two-stage attentive network (TSAN) for accurate SISR in this paper. As illustrated in Figure~\ref{fig:pipeline}, TSAN consists of two stages to solve the SISR problem in a coarse-to-fine manner. At LR-stage, we adopt a dilated residual block (DRB) as a fundamental unit to efficiently extract contextual features and further, based on DRB, propose a multi-context attentive block (MCAB) to make the network focus on more informative contextual features. 
Multiple MCABs are leveraged to extract attentive contextual features used for reconstructing an initial SR result. At HR-stage, we propose a refined attention block (RAB) to refine the initial SR result to a more fine-detailed one by exploring useful cues in HR space. 

Specifically, the DRB pushes the boundaries of conventional cascading and parallel strategies for feature extraction, which could simultaneously distill features with different receptive fields and different context characteristics. Dilated convolutions are adopted in DRB to further explore more contextual information through the larger receptive field. Based on the compact yet powerful DRB, the well-designed MCAB could distill attentive contextual features
by introducing the attention mechanism. MCAB contains two branches: a contextual feature extraction branch in which several DRBs are stacked in a dense connection manner to enhance the capability of the network to extract contextual features, and an attention branch that consists of a cutting-splicing block (CSB), a $1^{st}$-order attention triplet, and a $2^{nd}$-order attention triplet. 
The CSB is proposed to extract abundant structure cues and self-similarities in local and global regions simultaneously. 
The purpose of the $1^{st}$-order triplet and the $2^{nd}$-order triplet is to enhance the discriminative learning ability of the network through the interaction between spatial and channel dimensions.
Unlike DRB and MCAB, RAB is designed to explore available cues in HR space to reconstruct fine-detailed HR image. The intuition behind the RAB is that the information in LR space is limited, and we believe features extracted in HR space could benefit a better recovery of images details.
As shown in Figure~\ref{fig:teaser}, our TSAN obtains better visual quality and recovers more image details than other state-of-the-art SR methods.

To sum up, the main contributions of this paper are threefold: (1) we propose a two-stage TSAN which could address the SISR problem in a coarse-to-fine manner; (2) we design a novel multi-context attentive block (MCAB) with cross-dimension interaction; (3) our TSAN outperforms the state-of-the-art SISR methods in terms of accuracy and visual effects.

%% file: related_works.tex
\section{RELATED WORK}
The related work can be divided into two categories, {\em i.e.}, {\em single image super-resolution} and {\em attention mechanisms}.

\subsection{Single image super-resolution}
Single image super-resolution has been extensively studied in the past few decades. Numerous SISR methods have been proposed, ranging from early conventional methods~\cite{De1962Bicubic,duchon1979lanczos,dai2007soft,fattal2007image} and traditional learning-based methods~\cite{huang2015single,chang2004super,xu2018efficient,liu2016retrieval}, to recent deep learning-based methods.
In particular, deep learning-based methods have led to dramatic improvements in SISR due to the powerful representational capability of deep networks.
In this section, we mainly detail the most relevant deep learning-based SISR methods that can be categorized into three types, depending on how the network approaches the SR problem by either pre-upsampling, post-upsampling, or sampling.

For pre-upsampling based methods such as SRCNN~\cite{dong2016image}, VDSR~\cite{kim2016accurate}, DRCN~\cite{kim2016deeply}, DRRN~\cite{tai2017image} and MemNet~\cite{tai2017memnet}, the upsampling operator, {\em i.e.}, bicubic interpolation, often results in visible reconstruction artifacts.
Moreover, as these methods only learned the mapping in HR space, the raw features cannot be extracted from the original LR image to enhance the representational power of the network.

\zjq{Post-upsampling based methods like FSRCNN~\cite{dong2016accelerating}, IDN~\cite{hui2018fast}, SRResNet~\cite{ledig2017photo}, EDSR~\cite{lim2017enhanced}, MSRN~\cite{li2018multi}, RCAN~\cite{zhang2018image}, RDN~\cite{zhang2018residual}, CARN~\cite{ahn2018fast}, RNAN~\cite{zhang2019rnan}, OISR~\cite{he2019ode}, SAN~\cite{dai2019second}, DNCL~\cite{xie2018fast}, and IMSSRnet~\cite{lei2020deep}, directly extracted features from input LR images and then used the features merely extracted in LR space to construct the HR image by a transposed/sub-pixel convolution layer.}
However, as these methods focused on extracting features in LR space, a deeper or wider complex network ({\em e.g.}, EDSR~\cite{lim2017enhanced} and RDN~\cite{zhang2018residual}) was required to obtain sufficient information, for the purpose of reconstructing fine-detailed HR images.
Besides, the setting of one-step upsampling at the end of the network would also increase the difficulties of training large scaling factors.

Regarding sampling based methods~\cite{lai2017deep,haris2018deep,qiu2019embedded}, different sampling strategies were adopted in the network for some specific purposes.
For example, LapSRN~\cite{lai2017deep} progressively reconstructed the SR predictions to ease the difficulties of training for large scaling factors. Due to limited available features in the LR space, DBPN~\cite{haris2018deep} proposed an iterative up-and-down sampling approach that could obtain HR features in different depths for SR reconstruction.
It is worth mentioning that our proposed TSAN is sampling based. %can be seen as a mid-upsampling based method.
We leverage well-designed MCABs to efficiently extract abundant attentive contextual features with long-range dependencies from the input image in LR space. We also distill HR features from the initial coarse HR image through RAB, aiming at refining more local details.

\begin{figure*}[t]
	\centering

	\includegraphics[width = 1\linewidth]{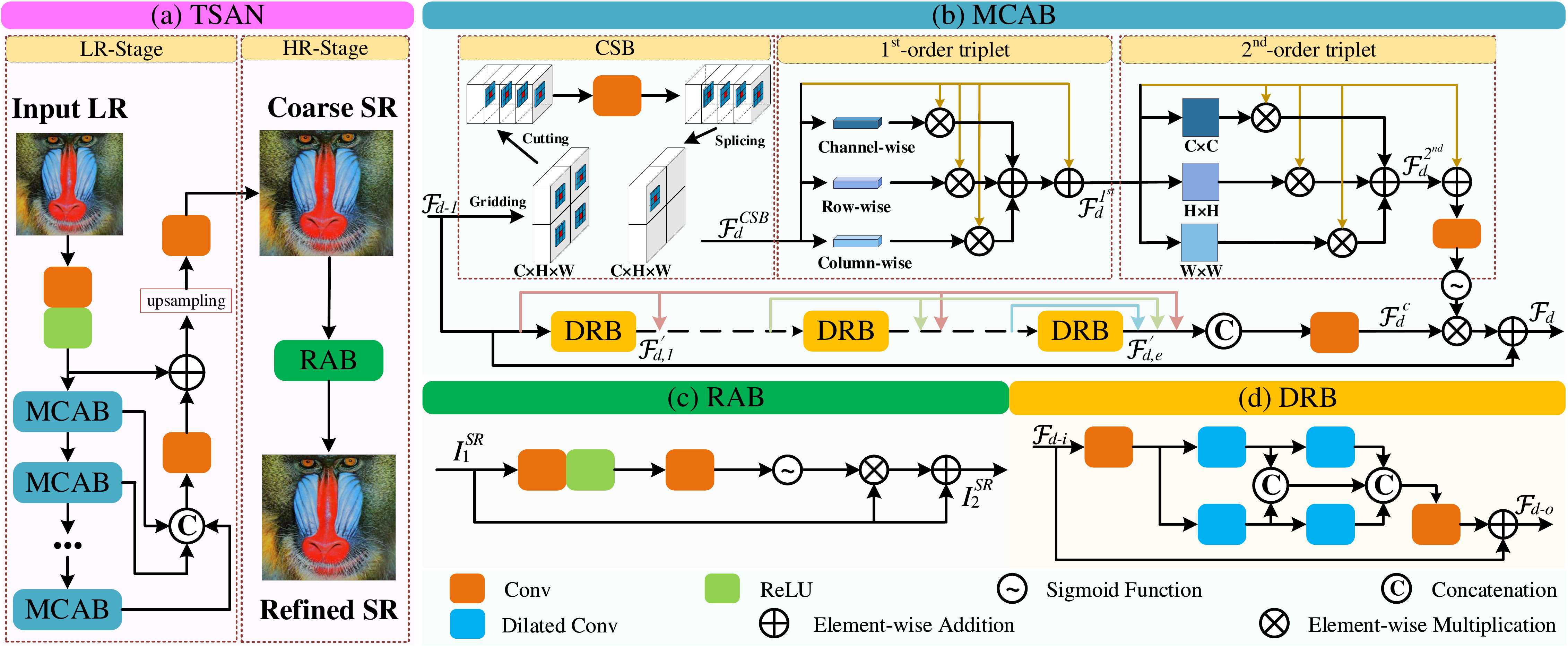}
	%  \vspace{-0.5cm}
	\caption{\zjq{(a) The overview of our proposed TSAN.  TSAN is a two-stage network which reconstructs SR image in a coarse-to-fine manner.  In LR-stage, MCABs are leveraged to extract attentive contextual features used for reconstructing an initial SR result. In HR-stage, RAB refines the initial SR result to a more fine-detailed one by exploring useful cues in HR space. (b) Our proposed Multi-Context Attentive Block (MCAB). (c) Our proposed Refined Attention Block (RAB).  (d) the Dilated Residual Block (DRB). }} 
	\label{fig:pipeline}
\end{figure*}

\subsection{Attention mechanisms}
Attention in human perception generally means that human visual systems adaptively process visual information and focus on salient areas. \zjq{Attention mechanisms have been widely applied in many tasks~\cite{zhao2017diversified,fu2019dual,Pramono_2019_ICCV,Chen_2019_ICCV}, including image super-resolution~\cite{zhang2018image, dai2019second,du2019orientation,hu2019channel,wu2020multi}}. Zhang et al.~\cite{zhang2018image} introduced attention mechanisms into the residual in residual structure to adaptively rescale channel-wise features for image super-resolution. Dai et al. ~\cite{dai2019second} proposed a second-order channel attention module to learn feature interdependencies by global covariance pooling for more discriminative representations.  
\zjq{Hu et al.~\cite{hu2019channel} constructed a set of channel-wise and spatial attention residual blocks and stacked them in a chain structure to dynamically modulate the multi-level features in global and local manners. Du et al.~\cite{du2019orientation} extracted orientation-aware features and combined them by a channel-wise attention mechanism to generate more distinctive features.
Wu et al.~\cite{wu2020multi} exploited the advantages of multi-scale and attention mechanisms in SR tasks.
However, the inter-dependence between the channel dimension and the spatial dimension is absent in the above-mentioned SISR methods when computing attention on these single pixel channels.  
Therefore, we propose the $1^{st}$-order and $2^{nd}$-order triplet attention to focus on inter-dependencies among channel dimension and different spatial dimensions.}

%% file: methodology.tex
\section{Methodology}

\subsection{\zjq{Overview}}
As illustrated in Figure~\ref{fig:pipeline}, our proposed TSAN consists of two stages to solve the SISR problem in a coarse-to-fine manner.
At LR-stage,  several multi-context attentive blocks (MCABs) are proposed to efficiently extract sufficient contextual features from the input LR image and construct an initial SR result based on the extracted features.
At HR-stage, a simple yet effective refined attention block (RAB) is proposed to further refine the coarse SR result obtained in LR-stage to a more fine-detailed one.

Given an input LR image $I^{LR}$, we first extract shallow features $\mathcal{F}_{s}$ by
\begin{equation}
\mathcal{F}_{s} = \delta(\mathcal{C}_{1 \times 1}(I^{LR})),
\end{equation}
where $\mathcal{C}_{k \times k}$ represents convolution operation where kernel size is $k \times k$; $\delta$ denotes the rectified linear unit (ReLU) activation function.
Then, $\mathcal{F}_{s}$ is fed to stacked multiple MCABs to distill attentive contextual features of different levels,
\begin{equation}
\begin{aligned}
\mathcal{F}_{d} = \mathcal{M}_{d}(\mathcal{F}_{d-1}) = \mathcal{M}_{d}(\mathcal{M}_{d-1}(\cdots \mathcal{M}_{1}(\mathcal{F}_s) \cdots)), 
\end{aligned}
\end{equation}
where $\mathcal{M}_{d}$ denotes the $d$-th MCAB and $\mathcal{F}_d$ denotes the output of the $d$-th MCAB. Here we embed three MCABs, \emph{i.e.}, $d=3$. Later, all $\mathcal{F}_i, i\in{[1, d]}$ are fused by applying a convolution layer upon the concatenation, {\em i.e.},
\begin{equation}
\mathcal{F}_{fusion} = \mathcal{C}_{1\times1}([\mathcal{F}_1, \mathcal{F}_2, \cdots, \mathcal{F}_d]),
\end{equation}
where $[\cdot]$ denotes the concatenation operation.
After that, we can get an initial SR result by applying a convolution layer upon the upsampled element-wise addition of the aggregated hierarchical features $\mathcal{F}_{fusion}$ and the raw features $\mathcal{F}_{s}$,
{\em i.e.},
\begin{equation}
I^{SR}_{1} = \mathcal{C}_{1\times1}(\kappa(\mathcal{F}_{fusion} + \mathcal{F}_{s})),
\end{equation}
where $\kappa$ denotes the sub-pixel operation.

Then we design a refined attention block (RAB) (denoted as $\mathcal{R}$) to refine the initial SR result by modeling the local details in HR space in the form of residual:
\begin{equation}
I_{2}^{SR} = \mathcal{R}(I^{SR}_{1}).
\end{equation}

Finally, these two stages are optimized jointly with the loss function defined as
\begin{equation}
\mathcal{L} = w_1\mathcal{L}_{m}(I_{1}^{SR}, I^{GT}) + w_2\mathcal{L}_{m}(I_{2}^{SR}, I^{GT}),
\label{equ:loss}
\end{equation}
where $I^{GT}$ is the ground truth image, $\mathcal{L}_{m}$ is the mean absolute error (MAE) loss, and $w_1$ and $w_2$ are the balancing parameters.

\subsection{\zjq{Multi-Context Attentive Block}}
As not all features contribute a positive effect to the desired SR result, we propose the multi-context attentive block (MCAB) to distill attentive contextual features with long-range dependencies for high-quality SR reconstruction.
Specifically, MCAB contains two branches: a contextual feature extraction branch (the lower part of Figure~\ref{fig:pipeline}(b)) and an attention branch (the upper part of Figure~\ref{fig:pipeline}(b)).

\subsubsection{\zjq{the contextual feature extraction branch}}
In the contextual feature extraction branch, the Dilated Residual Block (DRB) is adopted as the fundamental unit to explore more context cues by enlarging receptive field, and simultaneously extract features with different contextual characteristics for reconstructing visually pleasant HR image.

Cascading several convolution layers usually is an effective way to enlarge the receptive fields. In the cascading structure, as shown in Figure~\ref{fig:cp}(a), as a deeper layer accepts the output of a shallower layer, large receptive fields can be produced efficiently. Then the output of each layer would be fused to obtain features covering different scales of receptive fields. \zjq{Mathematically,
	\begin{equation}
	\begin{split}
	\mathcal{F}_{co} = & \mathcal{F}_{ci} + \mathcal{C}_{1\times1}([\mathcal{D}_{3\times3}^1(\mathcal{F}'), \mathcal{D}_{3\times3}^2(\mathcal{F}'), \\
	& \quad \quad \quad \quad   \quad \mathcal{D}_{3\times3}^3(\mathcal{F}'),\mathcal{D}_{3\times3}^4(\mathcal{F}')]), \\
	\mathcal{F}' = &  \mathcal{C}_{1\times1}(\mathcal{F}_{ci}),
	\end{split}	
	\end{equation}
where $\mathcal{F}_{ci}$ and $\mathcal{F}_{co}$ are the input and output of the cascaded network. $\mathcal{D}_{k \times k}^n$ represents dilated convolution operation where kernel size is $k \times k$, and n represents the number of consecutive use of the same dilated convolution on $\mathcal{F}'$.}
On the other hand, employing a parallel structure can harvest features with different context characteristics. In the parallel structure, as shown in Figure~\ref{fig:cp}(b), as multiple different convolution layers accept the same input and their outputs are concatenated together, the obtained output is indeed a sampling of the input using different contexts. \zjq{Mathematically,
	\begin{equation}
	\begin{split}
	\mathcal{F}_{po} = & \mathcal{F}_{pi} + \mathcal{C}_{1\times1}([\mathcal{D}_{3\times3}^1(\mathcal{F}'), \mathcal{D}_{3\times3}^1(\mathcal{F}'), \\
	& \quad \quad \quad \quad   \quad  \mathcal{D}_{3\times3}^1(\mathcal{F}'),\mathcal{D}_{3\times3}^1(\mathcal{F}')]), \\
	\mathcal{F}' = &  \mathcal{C}_{1\times1}(\mathcal{F}_{pi}),
	\end{split}	
	\end{equation}
	where  $\mathcal{F}_{pi}$ and $\mathcal{F}_{po}$ are the input and output of the parallel network.
}

To simultaneously distill features with different receptive fields and different context characteristics, we incorporate the advantages of both the cascading and parallel strategies and propose a novel compact structure shown in Figure~\ref{fig:pipeline}(d). The proposed DRB consists of two branches used for extracting features with different contextual characteristics. Each branch contains two cascaded convolution layers used for distilling features with different receptive fields. In order to obtain a larger receptive field without increasing the number of convolutions and parameters, we adopt dilated convolution. Finally, we concatenate all features of different branches and depths, and fuse them with input by a residual operation.
\zjq{For a clear presentation, DRB can be formulated as:
	\begin{equation}
	\begin{split}
	\mathcal{F}_{d-o} = & \mathcal{F}_{d-i} + \mathcal{C}_{1\times1}([\mathcal{D}_{3\times3}^1(\mathcal{F}'), \mathcal{D}_{3\times3}^1(\mathcal{F}'), \\
	& \quad \quad \quad \quad  \quad \quad \mathcal{D}_{3\times3}^2(\mathcal{F}'), \mathcal{D}_{3\times3}^2(\mathcal{F}')]), \\
	\mathcal{F}' = &  \mathcal{C}_{1\times1}(\mathcal{F}_{d-i}),
	\end{split}
	\end{equation}
	where $\mathcal{F}_{d-i}$ and $\mathcal{F}_{d-o}$ denote the input and output of DRB, respectively. The number of output channels for all dilated convolutionlayer in DRB is 64.}

\def\wdenoising{1\linewidth}
\def\hdenoising{0.45in}
\begin{figure}[t]
	\setlength{\tabcolsep}{2.4pt}
	\centering
	%	\vspace{-0.2cm}
	\begin{tabular}{c}
		\includegraphics[width=\wdenoising, height=\hdenoising]{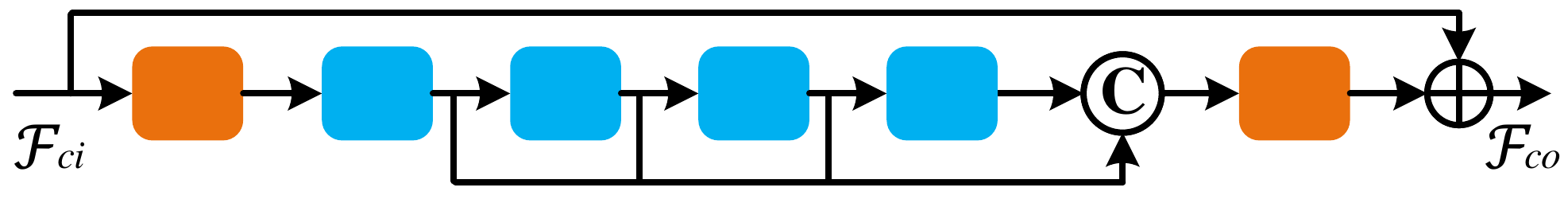} \\
		(a) \\
		\includegraphics[width=\wdenoising, height=\hdenoising]{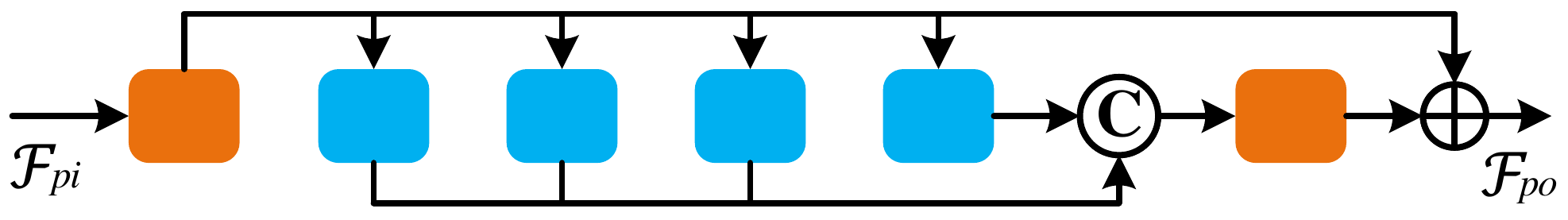} \\
		(b) \\
		
	\end{tabular}
	
	\caption{\zjq{(a) the cascaded structure; (b) the parallel structure.}}
	%	\vspace{-0.3cm}
	\label{fig:cp}
\end{figure}
%the core component, which  we use multiple DRBs to enhance the capability of the network to extract contextual features. 
%Dilated residual block (DRB) is proposed as the fundamental unit to explore more context cues by enlarging receptive field, and simultaneously extract features with different contextual characteristics for reconstructing visually pleasant HR image.

We further stack multiple DRBs in a dense connection manner, so that each DRB in MCAB has access to all the previous DRBs' output and could fully utilize them to further distill higher-level contextual features.
We then concatenate the outputs of each DRB and integrate them by a $1\times1$ convolution. The whole process can be expressed as:
\begin{equation}
\mathcal{F}_{d}^c= \mathcal{C}_{1 \times 1}([\mathcal{F}_{d-1}, \mathcal{F}_{d,1}^\prime, \cdots , \mathcal{F}_{d,e}^\prime]),
\end{equation}
where  $\mathcal{F}_{d}^c$ is the output of contextual feature extraction branch in $d$-th MCAB, $\mathcal{F}_{d, e}^\prime$ denotes the output of $e$-th DRB in $d$-th MCAB.
We use six DRBs in each MCAB, {\em i.e.}, $e=6$, and the dilation rates $s$ of these six DRBs are set to 1, 2, 3, 3, 2, and 1, respectively.

% $1^{st}$ and $2^{nd}$ order attention
\subsubsection{\zjq{the attention branch}}
The attention branch is designed to make the network focus on more informative features and enhance the discriminative learning ability of the network by considering long-range feature correlations in spatial and channel dimensions. As shown in Figure~\ref{fig:pipeline}(b), this branch contains three parts: a cutting-splicing block (CSB), a $1^{st}$-order triplet, and a $2^{nd}$-order triplet.

\textbf{\zjq{CSB.}} To simultaneously capture the spatial dependencies in the local patch and global diagram, we proposed a novel cutting-splicing block (CSB) to extract local patterns and exploit the abundant structure cues and self-similarities in global regions.
Formally, the features with a size of $C\times H\times W$ first be cut into $n\times n$ cells ($n=2$ in Figure~\ref{fig:pipeline}(b)), and then these cells are concatenated and fed into a $3\times 3$ convolution to extract and aggregate local and non-local information. After that, we splice $n \times n$  cells back into $C\times H\times W$ features. This process can be formulated as:
\begin{equation}
\mathcal{F}_{d}^{CSB} = \mathcal{O}_s(C_{3\times 3}(\mathcal{O}_c(\mathcal{F}_{d-1}))),
\end{equation} 
where $\mathcal{O}_s$ denotes splicing operation, $\mathcal{O}_c$ denotes cutting operation, and $\mathcal{F}_{d}^{CSB}$ is the output of CSB. The CSB is similar to introduce holes in dilated convolution. The difference is that we also consider the local neighbors. In this way, the local patterns and global diagram are simultaneously guaranteed.

\textbf{\zjq{$\bm{1^{st}}$-order triplet.}} After CSB, we design a $1^{st}$-order triplet and $2^{nd}$-order triplet to model inter-dependencies in different dimensions to find out regions/patterns that should be emphasized in contextual features. As the name implies, each attention triplet consists of three branches which are responsible for capturing cross-dimension interaction between the ($C$, $H$), ($C$, $W$), and ($H$, $W$) dimensions of the input tensor, respectively. 
By exploiting the inter-dependencies between the channel dimension and the spatial dimension, our network can effectively focus on informative contextual features.

In the $1^{st}$-order triplet, we take channel-wise attention as an example, we first aggregate spatial information of ($H$, $W$) dimension into a channel-wise descriptor by using average-pooling operation on each channel. Then, the descriptor is forwarded to a shared multi-layer perception ($\mathcal{MLP}$) to produce channel-wise attention maps,
\begin{equation}
\mathcal{A}_{cha}^{1^{st}} = \eta(\mathcal{MLP}(\psi(\mathcal{F}_{d}^{CSB}))),
\end{equation}
where $\eta$ is the sigmoid function, $\psi$ denotes global average pooling operation, $\mathcal{A}_{cha}^{1^{st}}$ denotes the channel-wise attention map. Then we can get channel-wise attentive features $\mathcal{F}_{cha}^{1^{st}}$ by multiplication between $\mathcal{F}_{d}^{CSB}$ and $\mathcal{A}_{cha}^{1^{st}}$. Similarly, we can obtain the row-wise attentive features $\mathcal{F}_{row}^{1^{st}}$ from ($C$, $H$) dimension and the column-wise attentive features $\mathcal{F}_{col}^{1^{st}}$ from ($C$, $W$) dimension. Then we add  $1^{st}$-order triplet attentive features and $\mathcal{F}_{d}^{CSB}$,
\begin{equation}
\mathcal{F}_{d}^{1^{st}} =  \mathcal{F}_{cha}^{1^{st}} +  \mathcal{F}_{row}^{1^{st}} + \mathcal{F}_{col}^{1^{st}} + \mathcal{F}_{d}^{CSB},
\end{equation}
where $\mathcal{F}_{d}^{1^{st}}$ denotes the output of the $1^{st}$-order triplet.

\textbf{\zjq{$\bm{2^{nd}}$-order triplet.}} Recent works~\cite{li2017second, dai2019second} have shown that second-order statistics in CNNs can provide different information for discriminative representations from the first-order ones.
Therefore, we also propose a $2^{nd}$-order triplet to learn feature inter-dependencies by cross-dimension interaction like in the $1^{st}$-order triplet.
In the $2^{nd}$-order triplet, we still use channel-wise attention (($H$, $W$) dimension) as an example. 
Specifically, we first apply average-pooling operation $\mathcal{M}$ along the channel axis on  $\mathcal{F}_{d}^{1^{st}} \in R^{C \times H \times W} $ to generate an efficient feature descriptor. Based on the feature descriptor, we apply a convolution layer and a sigmoid function to generate a spatial attention map $\mathcal{A}_{cha}^{2^{nd}}$ which encodes where to emphasize or suppress. 
Finally, we perform an multiplication operation between $\mathcal{A}_{cha}^{2^{nd}}$ and $\mathcal{F}_{d}^{1^{st}}$ to obtain spatial-wise attentive features $\mathcal{F}_{cha}^{2^{nd}}$. The whole process can be expressed as:
\begin{equation}
\begin{aligned}
\mathcal{F}_{cha}^{2^{nd}} = & \mathcal{A}_{cha}^{2^{nd}} \times \mathcal{F}_{d}^{1^{st}}, \\
\mathcal{A}_{cha}^{2^{nd}} = & \eta(\mathcal{C}_{1\times1}(\mathcal{M}(\mathcal{F}_{d}^{1^{st}}))),
\end{aligned}
\end{equation}
where $\eta$ is the sigmoid function.
Similarly, we generate the row-to-row features $\mathcal{F}_{row}^{2^{nd}}$ from ($C$, $H$) dimension and the column-to-column features $\mathcal{F}_{col}^{2^{nd}}$ from ($C$, $W$) dimension.
Then we add  $2^{nd}$-order triplet attentive features and $\mathcal{F}_{d}^{1^{st}}$,
\begin{equation}
\mathcal{F}_{d}^{2^{nd}} =  \mathcal{F}_{cha}^{2^{nd}} +  \mathcal{F}_{row}^{2^{nd}} +  \mathcal{F}_{col}^{2^{nd}} + \mathcal{F}_{d}^{1^{st}},
\end{equation}
Finally, we can obtain the output of $d$-th MCAB by,
\begin{equation}
\mathcal{F}_{d} = \mathcal{F}_{d}^c \times \eta(\mathcal{C}_{3\times3}(\mathcal{F}_{d}^{2^{nd}})) + \mathcal{F}_{d-1}.
\end{equation}
By capturing the inter-dependencies in different dimensions, the attention branch is able to focus on more informative features and enhance discriminative
learning ability. 
%channel-to-channel relationship as well as row-to-row and column-to-column relationships, the attention branch has a non-local contextual view and selectively aggregates features according to the attention map.
%

\subsection{\zjq{Refined Attention Block}}
As shown in Figure~\ref{fig:pipeline}(c), our refined attention block (RAB) is proposed to refine a coarse SR result to a more fine-detailed one. 
The RAB can be simply expressed as,
\begin{equation}
I_{2}^{SR} = I^{SR}_{1} \times \eta(\mathcal{C}_{1 \times 1}( \delta(\mathcal{C}_{3 \times 3}(I^{SR}_{1})))) +   I^{SR}_{1} ,
\end{equation}
Note that even though it looks very simple, RAB is essential for reconstructing a visually pleasant SR image with fine details. This is because the information in LR space is limited, and RAB can compensate for the lacked important local information by distilling features in HR space.

\subsection{\zjq{Implementation Details}}
We implement our model with Pytorch and run experiments with an NVIDIA Titan V GPU. For training, we use 48$\times$48 RGB patches cropped from LR image as input and its corresponding HR patches as ground truth. Following~\cite{lim2017enhanced}, we pre-process all the images by subtracting the mean RGB value of the DIV2K dataset~\cite{timofte2017ntire} and augment the training data with random horizontal flips and 90$^{\circ}$ rotations. We train our model with ADAM optimizer~\cite{kingma2015adam} by setting $\beta_1=0.9$, $\beta_2=0.999$. The mini-batch size is set to 16. The learning rate is initialized as 0.0001 and decreases to half every 200 epochs. And the number of total epochs is 1000. The balancing parameters $w_1$ and $w_2$ in Equation~\ref{equ:loss} are empirically set to 1.

\subsection{\zjq{Discussions}}
\subsubsection{Difference to RDN~\cite{zhang2018residual}} Inspired by RDN~\cite{zhang2018residual}, we introduce dense connection into our MCAB to fully utilize the features information from each DRB in MCAB.  There are some differences between RDN~\cite{zhang2018residual} and our TSAN.
First, RDN~\cite{zhang2018residual} mainly cascades convolution layers to enlarge the receptive field, while our TSAN is built based on DRB which can simultaneously distill features with different receptive fields and  different context characteristics. Second, RDN~\cite{zhang2018residual} focuses on how to exploit and use hierarchical features without considering how to distinguish different feature information. While our TSAN can learn the attention maps used for emphasizing informative contextual features by considering long-range feature correlations in spatial and channel dimensions. Third, RDN belongs to post-upsampling
methods which pay attention to exploiting the features information from LR space, while our TSAN is designed to utilize features from LR and HR space simultaneously. 

\subsubsection{Difference to RCAN~\cite{zhang2018image}} We summarize the main differences between RCAN~\cite{zhang2018image} and our TSAN. First, RCAN~\cite{zhang2018image} consists of several residual groups with long skip connections. While, TSAN stack DRBs in a dense connection manner, so each DRB in MCAB has access to all the previous DRBs’ output and could fully utilize them to further distill higher-level contextual features. Second, RCAN~\cite{zhang2018image} only extract local information for reconstructing. While TSAN considers non-local operations in CSB to learn long-range feature correlations. Third, RCAN~\cite{zhang2018image} only considers channel attention based first-order feature statistics to enhance the discriminative ability of the network. While our TSAN learns inter-dependencies between different dimensions based on first-order and second-order features.

\subsubsection{Difference to MSRN~\cite{li2018multi}} MSRN~\cite{li2018multi} proposes a multi-scale residual block (MSRB) to detect image features and fuse different scales. There are some main differences between MSRB in MSRN~\cite{li2018multi} and our proposed DRB. First, we utilize the dilated convolution to widen the receptive field without additive parameters, which maintains the lightweight structure of DRB. Second, MSRB is adopted to detect the image features at different scales, while we concatenate the outputs of four dilated convolution in DRB, which can simultaneously distill features with different receptive fields and  different scales features.

\subsubsection{Difference to MCERN~\cite{zhang2020multi}} MCERN~\cite{zhang2020multi} proposes a multi-context block and enhanced reconstruction network for SISR in a coarse-to-fine manner. There are some differences between MCERN~\cite{zhang2020multi} and our TSAN. First, MCERN~\cite{zhang2020multi} only focuses on how to extract rich contextual information, while our TSAN  considers informative features based on MCERN~\cite{zhang2020multi}. 
Second, MCERN~\cite{zhang2020multi} only utilize local information for SISR, while TSAN can guarantee local patterns and global diagram with CSB.

\subsubsection{Difference to SAN~\cite{dai2019second}} SAN~\cite{dai2019second} introduces second-order attention operations to learn feature inter-dependencies by global covariance pooling for more discriminative representations in image super-resolution. The main differences between SAN~\cite{dai2019second} and our TSAN lie in the following aspects. First, SAN~\cite{dai2019second} pays attention to make full use of
the information from the original LR images, while our TSAN values the features information of the LR space and the HR space, and processes the obtained features information in a coarse to fine manner. Second, SAN~\cite{dai2019second} presents a non-locally enhanced residual group structure based on~\cite{wang2018non} to capture long-distance contextual information. While we propose a simple and flexible CSB to exploit long-range feature correlation, and we use the triplet structure to capture cross-dimension interaction between the ($C$, $H$), ($C$, $W$), and ($H$, $W$) dimensions.

%% file: experiments.tex
\section{EXPERIMENT}

To verify the effectiveness of the proposed method, we evaluate our SR results with two metrics, {\em i.e.}, peak signal-to-noise ratio (PSNR) (unit: dB) and structural similarity (SSIM)~\cite{wang2004image} on Y channel of transformed YCbCr space. For the convenience of a fair comparison, we follow the
experiment setup of existing methods. To specify, we use the high-quality DIV2K~\cite{timofte2017ntire} dataset for training and take four test sets - Set5~\cite{bevilacqua2012low}, Set14~\cite{zeyde2010single}, BSDS100~\cite{arbelaez2011contour}, and Urban100~\cite{huang2015single} for evaluation.

\def\wdenoising{0.315\linewidth}
\def\hdenoising{0.7in}
\begin{figure}[b]
	\setlength{\tabcolsep}{2.4pt}
	\centering
	%	\vspace{-0.2cm}
	\begin{tabular}{ccc}
		\includegraphics[width=\wdenoising, height=\hdenoising]{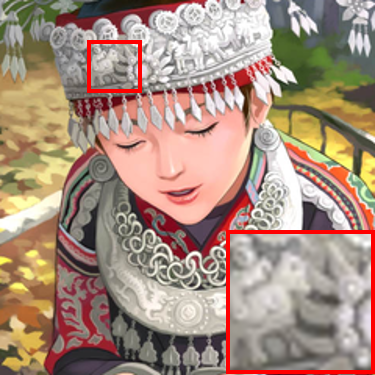}&
		\includegraphics[width=\wdenoising, height=\hdenoising]{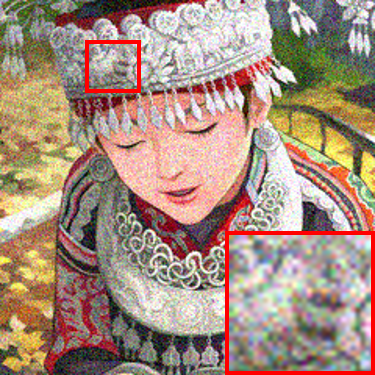}&
		\includegraphics[width=\wdenoising, height=\hdenoising]{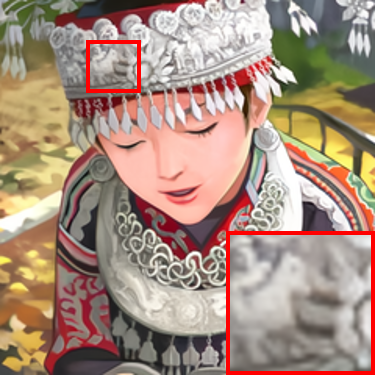}\\
		
		\includegraphics[width=\wdenoising, height=\hdenoising]{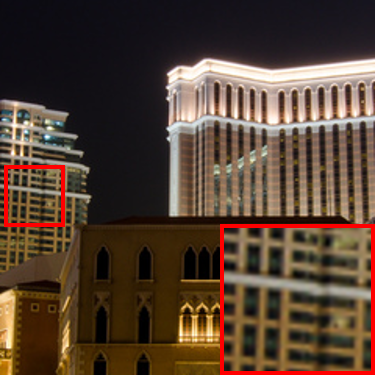}&
		\includegraphics[width=\wdenoising, height=\hdenoising]{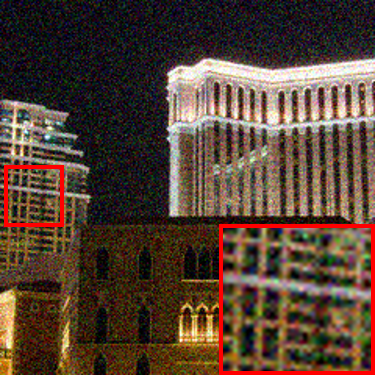}&
		\includegraphics[width=\wdenoising, height=\hdenoising]{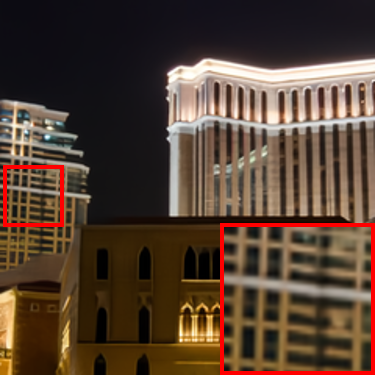}\\
		
		(a) Original & (b) Noisy & (c) TSAN\\
		
	\end{tabular}
	
	\caption{\zjq{Qualitative comparisons of color image denoising. The second column shows the noisy images with noise level 25. TSAN recovers fine local details, which is mainly contributed by the abundant hierarchical and contextual features extracted by our proposed DRB. Best viewed in zoom in.}}
	%	\vspace{-0.3cm}
	\label{fig:denoising}
\end{figure}

\setlength{\tabcolsep}{2.8pt}
\begin{table}[t]
	\centering
	%	\small
	\caption{DRB architecture analysis with $\times$2 scale factor. ``Cascaded'' and ``Parallel''  denote the cascaded architecture and the parallel architecture, respectively. $s=1$ means the dilation rate of all dilated convolutions in the DRB is 1.}
	\label{tab:multibranch}
	\begin{tabular}{lcccc}%{p{1.2cm}<{\centering} p{0.8cm}<{\centering} p{0.8cm}<{\centering} p{0.8cm}<{\centering} p{0.8cm}<{\centering} p{0.8cm}<{\centering} p{0.8cm}<{\centering}}
		\hline
		\multirow{2}{*}{} &  Set5~\cite{bevilacqua2012low} &  Set14~\cite{zeyde2010single} &BSDS100~\cite{arbelaez2011contour} &Urban100~\cite{huang2015single} \\
		& PSNR/SSIM  & PSNR/SSIM  & PSNR/SSIM & PSNR/SSIM \\
		\hline
		Cascaded  &38.14/0.9611  & 33.78/0.9191  & 32.27/0.9009  & 32.50/0.9322  \\
		Parallel  & 38.10/0.9609   & 33.79/0.9189  & 32.25/0.9004   & 32.44/0.9311       \\
		DRB(s = 1)  & 38.17/0.9609   & 33.80/0.9186  & 32.27/0.9006 & 32.54/0.9328     \\
		DRB   & \bf{38.22}/\bf{0.9613}  & \textbf{33.84}/\textbf{0.9196}   & \bf{32.32}/\bf{0.9015}   & \bf{32.77}/\bf{0.9345}  \\
		\hline
	\end{tabular}
	%  \vspace{-0.5cm}
\end{table}

\subsection{\zjq{Effectiveness of Key Components in MCAB}}
\zjq{MCAB contains two branches, \textit{i.e.}, the contextual feature extraction branch and the attention branch.}

\subsubsection{\zjq{Effectiveness of the Contextual Feature Extraction Branch}}
The core of this branch is DRB.
%\subsection{\zjq{Effectiveness of DRB}}
To verify the effectiveness of the DRB structure, we make three sets of experiments. In the first set of experiments, we replace our DRB structure with the cascaded structure (see Figure~\ref{fig:cp}(a)) or the parallel structure (see Figure~\ref{fig:cp}(b)) and evaluate their performances on the four datasets. For a fair comparison, we set $d=3$, $e=6$, and $s=1$ (that means the dilation rate of all dilated convolutions in the DRB is 1, there are 6 DRBs in each MCAB, and  there are 3 MCABs in the whole network). Note that the numbers of these three network parameters are the same. The results are summarized in Table~\ref{tab:multibranch}, from which we can claim that the proposed DRB structure performs best under the same parameters. Such an observation demonstrates that the multiple branches structure in the DRB is more efficient than cascading and parallel structures. This is because DRB can incorporate the advantages of both the cascading and parallel strategies to simultaneously distill features with different receptive fields and different context characteristics.

To demonstrate the importance of different dilated convolution with the dilation rate, we conduct a second set of experiments to compare with a variant structure in which we set the dilation rate $s$ of all dilated convolutions in the DRB to 1 (DRB(s=1)).  Similarly, we set $d=3$ and $e=6$. The PSNR and SSIM in Table~\ref{tab:multibranch} show that our setting of dilation rate achieves better results on test datasets. This suggests applying different dilated convolution with the dilation rate can obtain a larger receptive field for SISR.

In order to further verify the validity of our proposed DRB structure, we use our network at LR-stage for other low-level computer vision tasks. We provide the results of image denoising in Figure~\ref{fig:denoising}. Apparently, our proposed TSAN produces a good result on image
denoising because our DRB structure is able to extract abundant hierarchical and contextual features for image reconstruction.

\setlength{\tabcolsep}{1.0pt}
\begin{table}[bp]
	\centering
	\caption{\zjq{The ablation study of MCAB components. The results are evaluated on Set5~\cite{bevilacqua2012low} for a scale factor of $\times$3.}}
	\label{tab:ablation_att}
	\begin{tabular}{c||c||c|c||c|c||c}
		\hline
		& \multirow{2}{*}{w/o CSB}     & w/o $1^{st}$-     & $1^{st}$-order      & w/o $2^{nd}$-    & $2^{nd}$-order   &  \multirow{2}{*}{TSAN}   \\ 
		& & order triplet  & triplet\_HW & order triplet  &triplet\_HW & \\
		\hline
		PSNR &34.55 & 34.57& 34.60 &34.51 &34.57 & 34.64 \\
		\hline
	\end{tabular}
\end{table}

The above three experiments demonstrate that our proposed DRB is an effective structure that can distill features with different contextual characteristics by two branches and extract features in different receptive fields by two
cascaded convolution layers of each branch. And multiple DRBs are combined to integrate contextual features of different levels adaptively.

\subsubsection{\zjq{Effectiveness of the Attention Branch}} the attention branch consists of three important components, including CSB, $1^{st}$-order triplet, and $2^{nd}$-order triplet. To verify the effectiveness of different components, we compare TSAN without using CSB, $1^{st}$-order triplet, and $2^{nd}$-order triplet in Table~\ref{tab:ablation_att}. Further, to demonstrate the importance of inter-dependencies between the channel dimension and the spatial dimension, we remove cross-dimension interaction between the ($C$, $H$) and  ($C$, $W$) dimensions in $1^{st}$-order triplet and $2^{nd}$-order triplet, and only retain the interaction of ($H$, $W$) dimension, which denoted as $1^{st}$-order triplet\_HW, and $2^{nd}$-order triplet\_HW.
%its variants tested on Set5~\cite{bevilacqua2012low} dataset. The specific performance is listed in Table~\ref{tab:ablation_att}.
It can be found that the CSB contributes to performance improvement. This is mainly because CSB provides local and non-local information to the network, capturing short-distance and long-distance features simultaneously. We can also learn that $1^{st}$-order and $2^{nd}$-order triplet components contribute to the network ability obviously. This indicates discriminative learning with cross-dimension interaction plays an important role in determining the performance.

\subsection{\zjq{Effectiveness with Different Number of MCABs}}
%As we all know, increasing depth will promote the performance of the network.
We set different numbers of MCABs in our proposed TSAN and evaluate the performances on different datasets. As shown in Table~\ref{tab:mcbs}, the values of both PSNR and SSIM for our network get better as the number of MCABs increases. Such an observation is consistent with what we expect since the generalization ability will also increase when the number of parameters of our network will go up. As a trade-off between the performance and the complexity of the network,  we determine to use three MCABs, which provides strong reconstruction ability and requires not many parameters ($<$ 5.0M).

\setlength{\tabcolsep}{3.8pt}
\begin{table}[bp]
	\centering
	%	\small
	\caption{\zjq{MCABs analysis by varying the number of MCABs in TSAN. The scale factor is $\times$2.}}
	\label{tab:mcbs}
	\begin{tabular}{ccccc}
		\hline
		number of  & Set5~\cite{bevilacqua2012low} &Set14~\cite{zeyde2010single} &BSDS100~\cite{arbelaez2011contour} & Urban100~\cite{huang2015single} \\
		MCABs & PSNR/SSIM & PSNR/SSIM  & PSNR/SSIM & PSNR/SSIM \\
		\hline
		1  & 38.01/0.9604 & 33.55/0.9168 & 32.13/0.8992 & 31.92/0.9261    \\
		2  & 38.12/0.9609 & 33.74/0.9188 & 32.23/0.9003 & 32.27/0.9296    \\
		3  & 38.22/0.9613 & 33.84/0.9196  & 32.32/0.9015  & 32.77/0.9345     \\
		4  & \textbf{38.23}/\textbf{0.9614} & \bf{33.87}/\bf{0.9198}     & \bf{32.35}/\bf{0.9016}    & \bf{32.82}/\bf{0.9348}     \\
		\hline
	\end{tabular}
	%	\vspace{-0.2cm}
\end{table}

\subsection{\zjq{Effectiveness of RAB}}
To verify that RAB can further improve the reconstruction effect, we conduct a group of experiments without RAB. For a fair comparison, we move the RAB to the front of the upsampling to ensure that the depth of TSAN w/o and w/ RAB are the same. The results are summarized in Table~\ref{tab:rab}. From the results, we can clearly observe that the performance with RAB works better than that without RAB, which suggests that our RAB is able to refine a coarse HR result to a more detailed one since it can continue to extract useful features from HR space. We also visualize the visual effects of TSAN w/o and w/ RAB in Figure~\ref{fig:worefine}. In this example, our RAB is able to correct the direction for black lines. 
In order to further prove the effectiveness of our coarse-to-fine method, we designed a variant w/ RAB$^\dag$  that sets the parameters $w_1$  in the objective function to 0 and $w_2$ to 1. From Table~\ref{tab:rab}, it can be found that  $w_1$  equal to zero will have a negative impact on the reconstruction results. The reason behind this is that the lack of intermediate coarse results reconstructed from the LR space makes the training process difficult. This demonstrates that  multiple MCABs can extract rich attentive contextual features with cross-dimension interaction to obtain a good initial coarse SR result, then our proposed RAB can further improve the coarse SR image into a fine SR image by using LR and HR space information simultaneously.

\setlength{\tabcolsep}{2.5pt}
\begin{table}[tb]
	\centering
	%	\small
	\caption{\zjq{The effectiveness of RAB. For a fair comparison, we move the RAB to the front of the upsampling to ensure that the depth of TSAN w/o and w/ RAB are the same. The results evaluated for a scale factor of $\times$2.}}
	\label{tab:rab}
	\begin{tabular}{lcccc}
		\hline
		\multirow{2}{*}{TSAN} & Set5~\cite{bevilacqua2012low} &Set14~\cite{zeyde2010single} &BSDS100~\cite{arbelaez2011contour} &Urban100~\cite{huang2015single} \\
		& PSNR/SSIM  & PSNR/SSIM & PSNR/SSIM & PSNR/SSIM \\
		\hline
		w/o RAB  & 38.14/0.9611    & 33.75/0.9192   & 32.29/0.9010 & 32.58/0.9327   \\
		w/ RAB$^\dag$  & 38.10/0.9610     & 33.79/0.9193  & 32.26/0.9007 & 32.49/0.9320 \\
		
		w/ RAB  & \bf{38.22}/\bf{0.9613}     & \bf{33.84}/\bf{0.9196}  & \bf{32.32}/\bf{0.9015} & \textbf{32.77}/\textbf{0.9345} \\
		
		\hline
	\end{tabular}
	%\vspace{-0.5cm}
\end{table}

\def\wdenoisinglarge{1\linewidth}
\def\hdenoisinglarge{1.2in}
\def\wdenoising{0.315\linewidth}
\def\hdenoising{0.7in}
\begin{figure}[bp]
	\setlength{\tabcolsep}{2.4pt}
	\centering
	%	\vspace{-0.2cm}
	\begin{tabular}{ccc}
		%    \multicolumn{3}{c}{\includegraphics[width=\wdenoisinglarge, height=\hdenoisinglarge]{figures/visual/worefine/Urban100-062/image.png}} \\
		\includegraphics[width=\wdenoising, height=\hdenoising]{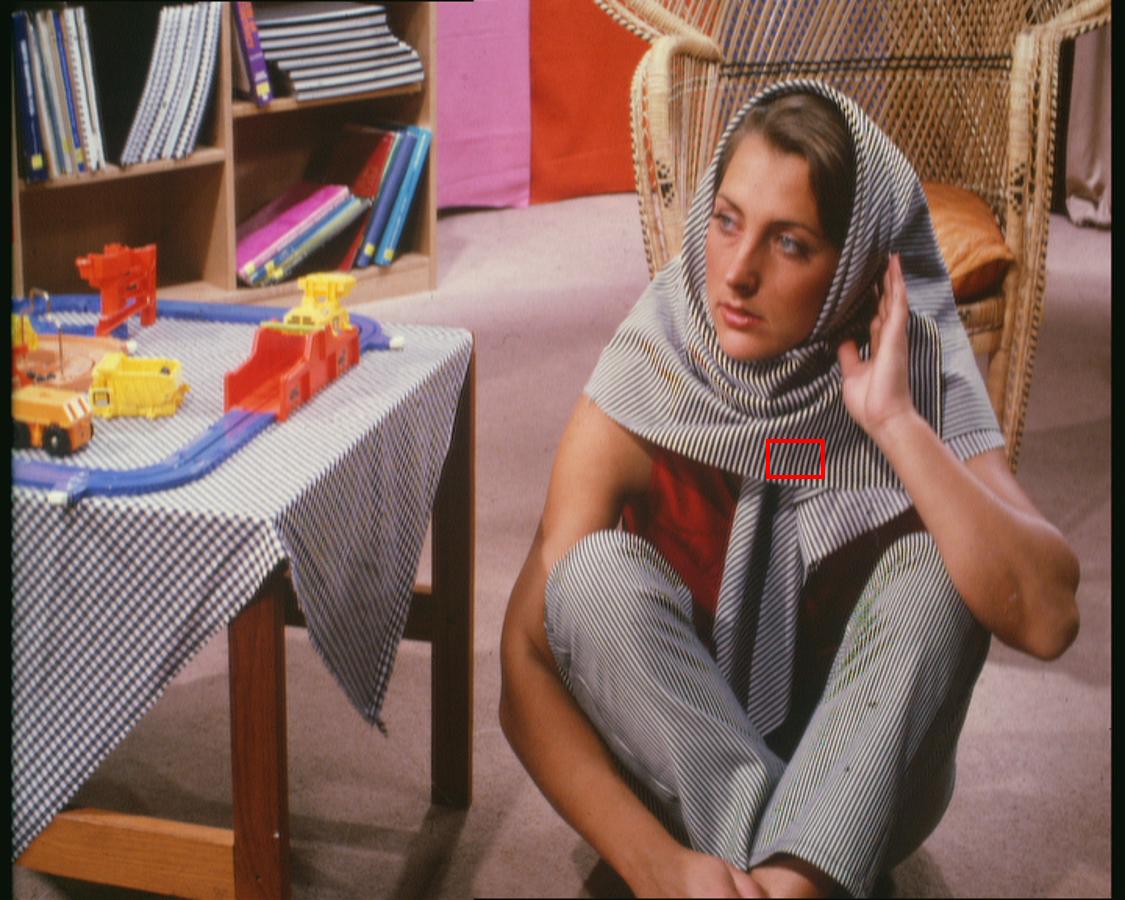}&
		\includegraphics[width=\wdenoising, height=\hdenoising]{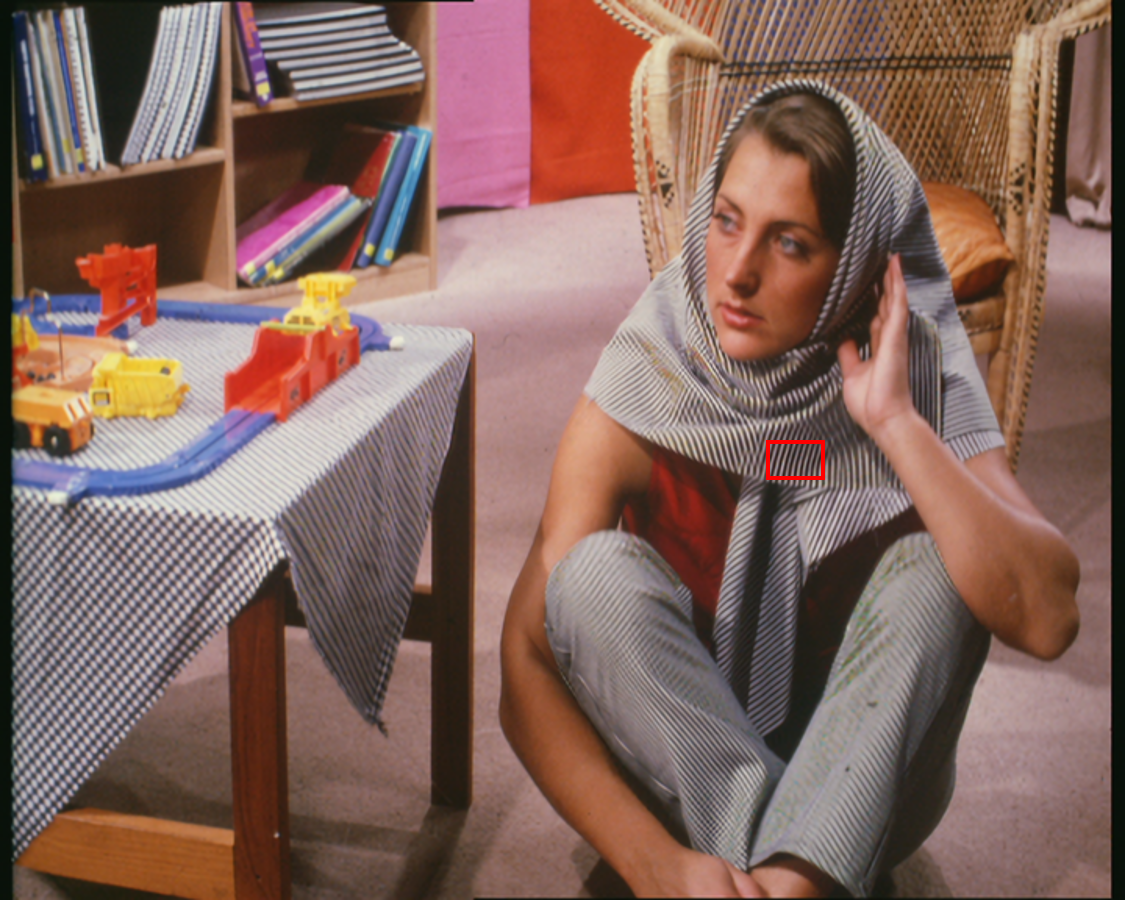}&
		\includegraphics[width=\wdenoising, height=\hdenoising]{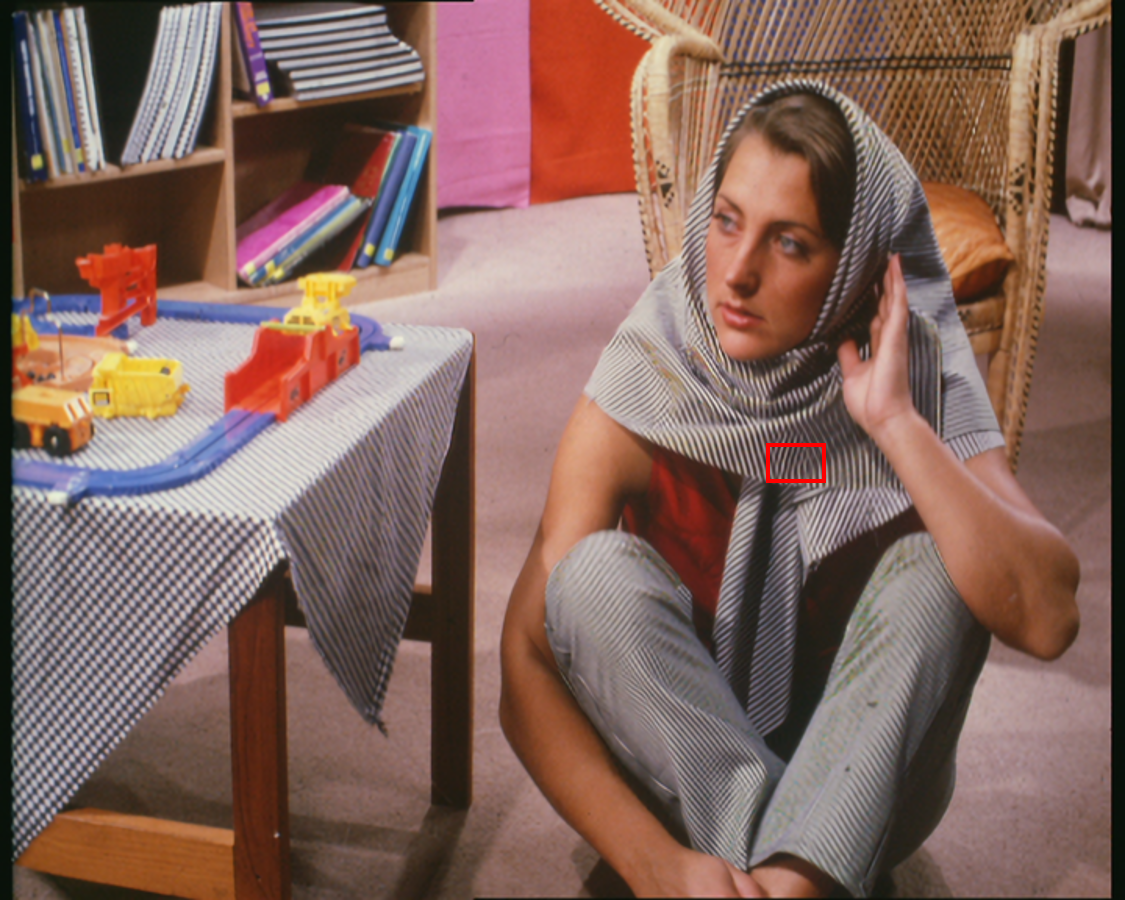}\\
		
		\includegraphics[width=\wdenoising, height=\hdenoising]{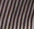}&
		\includegraphics[width=\wdenoising, height=\hdenoising]{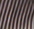}&
		\includegraphics[width=\wdenoising, height=\hdenoising]{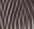}\\
		
		(a) GT  & (b) w/ RAB & (c) w/o RAB \\
		
	\end{tabular}
	%	\vspace{-0.2cm}
	\caption{\zjq{Visual comparisons of the effectiveness of RAB on \emph{barbara} from Set14~\cite{zeyde2010single} with scale factor $\times$2.}}
	\label{fig:worefine}
	%	\vspace{-0.3cm}
\end{figure}

\setlength{\tabcolsep}{17pt}
\begin{table*}[htbp]
	%	\small
	%	\large
	%\fontsize{14pt}{14pt}
	\centering
	\caption{\zjq{Performance comparison to other 8 state-of-the-art methods with the light-weighted model ($<$ 6.0M parameters). The best and second best
			results are highlighted in \textbf{bold} and \underline{underlined}.}}
	\label{tab:par}
	
	\begin{tabular}{c|l|c|c|c|c}%{p{1.4cm}<{\centering} p{1.2cm}<{\centering} p{1.5cm}<{\centering} p{0.7cm}<{\centering}  p{0.7cm}<{\centering} p{0.7cm}<{\centering}}
		\hline
		%		\specialrule{0em}{1.5pt}{1.5pt}
		\hline
		\multirow{2}{*}{Scale} & \multirow{2}{*}{Method} & Set5~\cite{bevilacqua2012low} & Set14~\cite{zeyde2010single} & BSDS100~\cite{arbelaez2011contour} & Urban100~\cite{huang2015single} \\
		%		\cline{3-4} \cline{5-6} \cline{7-8} \cline{9-10}
		&  & PSNR/SSIM   & PSNR/SSIM & PSNR/SSIM & PSNR/SSIM  \\
		\hline
		%		\specialrule{0em}{1.5pt}{1.5pt}
		\hline
		
		\multirow{9}{*}{$\times$2} & Bicubic & 33.66/0.9299& 30.24/0.8688  & 29.56/0.8431 & 26.88/0.8403\\
		& LapSRN~\cite{lai2017deep}  (CVPR'17) &  37.52/0.9591  & 33.08/0.9130 & 31.80/0.8950 & 30.41/0.9101  \\
		& CARN~\cite{ahn2018fast}  (ECCV'18) & 37.76/0.9590 & 33.52/0.9166 & 32.09/0.8978  & 31.51/0.9312  \\
		& SRMDNF~\cite{zhang2018learning}  (CVPR'18) & 37.79/0.9601 & 33.32/0.9159  & 32.05/0.8985  & 31.33/0.9204\\
		& NLRN~\cite{liu2018non}  (NIPS'18) &  38.00/0.9603 & 33.46/\underline{0.9195} & 32.19/0.8992  & 31.82/0.9249 \\
		& MSRN~\cite{li2018multi}  (ECCV'18) & 38.08/0.9605  & 33.74/0.9170 & 32.23/\underline{0.9013} & 32.22/\underline{0.9326} \\
		& FRSR~\cite{Soh_2019_CVPR} (CVPR'19) & 37.95v0.9594 & 33.45/\underline{0.9195} & 32.17/0.8991 & 32.23/0.9290 \\
		& OISR-RK2~\cite{he2019ode}  (CVPR'19)  & 38.11/0.9609 & \underline{33.80}/0.9193 & \underline{32.26}/0.9006 & \underline{32.48}/0.9317	  \\
		& LattienNet~\cite{luolatticenet} (ECCV'20)  & \underline{38.15}/\underline{0.9610}& 33.78/0.9193 & 32.25/0.9005 & 32.43/0.9302 \\
		%		& FSRCNN~\cite{lee2020learning} (ECCV'20) & &  & &  & & & &	 \\
		& TSAN & \textbf{38.22}/\textbf{0.9613} & \textbf{33.84}/\textbf{0.9196} & \textbf{32.32}/\textbf{0.9015}  & \textbf{32.77}/\textbf{0.9345} \\
		
		\hline
		\multirow{9}{*}{$\times$3} & Bicubic & 30.39/0.8682 & 27.55/0.7742 & 27.21/0.7385 & 24.46/0.7349 \\
		& LapSRN~\cite{lai2017deep}  (CVPR'17) & 33.82/0.9227 & 29.87/0.8320 & 28.82/0.7980 & 27.07/0.8280  \\
		& CARN~\cite{ahn2018fast}  (ECCV'18)  & 34.29/0.9255 & 30.29/0.8407 & 29.06/0.8034 & 27.38/0.8404 \\
		& SRMDNF~\cite{zhang2018learning}  (CVPR'18) & 34.12/0.9254  & 30.04/0.8382 & 28.97/0.8025 & 27.57/0.8398 \\
		& NLRN~\cite{liu2018non}  (NIPS'18) &  34.27/0.9266  & 30.16/0.8374 & 29.06/0.8026 & 27.93/0.8453  \\
		& MSRN~\cite{li2018multi}  (ECCV'18) &34.38/0.9262& 30.34/0.8395 & 29.08/0.8041 & 28.08/0.8554  \\
		& FRSR~\cite{Soh_2019_CVPR} (CVPR'19)  & 34.38/0.9262 & 30.27/0.8411 & 29.11/0.8050 & 28.33/0.8556  \\
		& OISR-RK2~\cite{he2019ode}  (CVPR'19)  & \underline{34.55}/\underline{0.9281} & \underline{30.46}/\underline{0.8443}  & \underline{29.18}/\underline{0.8075} & \underline{28.50}/\underline{0.8597}  \\
		& LattienNet~\cite{luolatticenet} (ECCV'20)  & 34.53/\underline{0.9281} &  30.39/0.8424 &  29.15/0.8059 &  28.33/0.8538 \\
		%		& FSRCNN~\cite{lee2020learning} (ECCV'20) & &  & &  & & & &	 \\
		& TSAN & \textbf{34.64}/\textbf{0.9282} &\textbf{30.52}/\textbf{0.8454} & \textbf{29.20}/\textbf{0.8080}  & \textbf{28.55}/\textbf{0.8602} \\
		
		\hline
		\multirow{9}{*}{$\times$4} & Bicubic &  28.42/0.8104 & 26.00/0.7027 & 25.96/0.6675 & 23.14/0.6577 \\
		& LapSRN~\cite{lai2017deep}  (CVPR'17) & 31.54/0.8850 & 28.19/0.7720 & 27.32/0.7270 & 25.21/0.7560  \\
		& CARN~\cite{ahn2018fast}  (ECCV'18) & 31.92/0.8903  & 28.42/0.7762 & 27.44/0.7304 & 25.63/0.7688 \\
		& SRMDNF~\cite{zhang2018learning}  (CVPR'18) & 31.96/0.8925& 28.35/0.7787 & 27.49/0.7337 & 25.68/0.7731  \\
		& NLRN~\cite{liu2018non}  (NIPS'18) & 31.92/0.8916 & 28.36/0.7745 & 27.48/0.7306 & 25.79/0.7729 \\
		& MSRN~\cite{li2018multi}  (ECCV'18)  & 32.07/0.8903 & 28.60/0.7751 & 27.52/0.7273 & 26.04/0.7896 \\
		& FRSR~\cite{Soh_2019_CVPR} (CVPR'19) &  32.22/0.8950 & 28.64/0.7830 & 27.60/0.7370 & 26.21/0.7910 \\
		& OISR-RK2~\cite{he2019ode}  (CVPR'19)   &\underline{32.35}/\underline{0.8970}  & \underline{28.72}/\underline{0.7843} & \underline{27.66}/\underline{0.7390} & \underline{26.37}/\underline{0.7953}    \\
		& LattienNet~\cite{luolatticenet} (ECCV'20)  & 32.30/0.8962 & 28.68/0.7830 & 27.62/0.7367 & 26.25/0.7873 \\
		%		& FSRCNN~\cite{lee2020learning} (ECCV'20) & &  & &  & & & &	 \\
		& TSAN &\textbf{32.40}/\textbf{0.8975} & \textbf{28.73}/\textbf{0.7847} & \textbf{27.67}/\textbf{0.7398}  &\textbf{26.39}/\textbf{0.7955} \\
		
		\hline
		%		\specialrule{0em}{1.5pt}{1.5pt}
		\hline	
	\end{tabular}
	%  \vspace{-0.5cm}
\end{table*}  

\setlength{\tabcolsep}{17pt}
\begin{table*}[htbp]
	%	\small
%	\large
	\centering
	%\caption{Comparison with MRSN and another two current state-of-the-art methods. The Multi-Adds is computed by assuming that the size of LR image is $ \bf{48\times48} $ and the scale factor is 2. ``avgPSNR" and ``avgSSIM" denote the average performance evaluated on four test sets.}
	\caption{\zjq{Performance comparison to 8 state-of-the-art methods with the heavy-weighted model. The best and second best
		results are highlighted in \textbf{bold} and \underline{underlined}.}}
	\label{tab:params}
	\begin{tabular}{c|l|c|c|c|c}%{p{1.4cm}<{\centering} p{1.2cm}<{\centering} p{1.5cm}<{\centering} p{0.7cm}<{\centering}  p{0.7cm}<{\centering} p{0.7cm}<{\centering}}
		\hline
		\hline
		
		\multirow{2}{*}{Scales} & \multirow{2}{*}{Methods}  &Set5~\cite{bevilacqua2012low} &Set14~\cite{zeyde2010single} &BSDS100~\cite{arbelaez2011contour} &Urban100~\cite{huang2015single}\\
		& & PSNR/SSIM & PSNR/SSIM & PSNR/SSIM & PSNR/SSIM \\
		
		\hline
		\hline
		\multirow{10}{*}{$\times$2} & EDSR~\cite{lim2017enhanced} (CVPRW'17)  & 38.11/0.9602  & 33.92/0.9195 & 32.32/0.9013 & 32.93/0.9351  \\
		& RDN~\cite{zhang2018residual} (CVPR'18)   & 38.24/0.9614  & 34.01/0.9212 & 32.34/0.9017  & 32.89/0.9353 \\
		& DBPN~\cite{haris2018deep} (CVPR'18)  & 38.09/0.9600 & 33.85/0.9190 & 32.27/0.9000& 32.55/0.9324 \\
		& RCAN~\cite{zhang2018image} (ECCV'18)  & 38.27/0.9614  & 34.12/0.9216  & \underline{32.40}/0.9025 & 33.34/0.9384 \\
		& RNAN~\cite{zhang2019residual} (ICLR'19)  & 38.17/0.9611 & 33.87/0.9207  & 32.32/0.9014&  32.73/0.9340 \\
		& SAN~\cite{dai2019second} (CVPR'19)   & \underline{38.28}/\underline{0.9618}  & 34.07/0.9213 & 32.35/0.9019& 33.10/0.9370  \\
		%  DBPN~\cite{haris2018deep} & 6.0($\times$1.6) &34.7($\times$3.4) & 32.63 & 0.9051 \\
		& Pan~\cite{pan2020physics_sr} (AAAI'20)   &38.26/0.9614   & 33.99/0.9200   &32.37/0.9020  &33.09/0.9365  \\
		& HAN~\cite{niu2020single} (ECCV'20)  & 38.27/0.9614 &\underline{34.16}/\underline{0.9217}   &\textbf{32.41/0.9027}   &\underline{33.35}/\underline{0.9385}  \\
		& \textbf{TSAN} &  38.22/0.9613 & 33.84/0.9196  & 32.32/0.9015 & 32.77/0.9345\\
		& \textbf{TSAN-L} & \textbf{38.30/0.9619} &\textbf{34.17/0.9218} &\underline{32.40}/\underline{0.9026}& \textbf{33.45/0.9387} \\
		
		\hline
		\multirow{10}{*}{$\times$3} & EDSR~\cite{lim2017enhanced} (CVPRW'17)  & 34.65/0.9280  & 30.52/0.8462 & 29.25/0.8093 & 28.80/0.8653  \\
		& RDN~\cite{zhang2018residual} (CVPR'18)   & 34.71/0.9296  & 30.57/0.8468 & 29.26/0.8093  & 28.80/0.8653 \\
		& DBPN~\cite{haris2018deep} (CVPR'18)  & --/-- & --/--  & --/--  & --/--  \\
		& RCAN~\cite{zhang2018image} (ECCV'18)  & 34.74/0.9299  & \underline{30.65}/0.8482  & 29.32/0.8111 & 29.09/0.8702 \\
		& RNAN~\cite{zhang2019residual} (ICLR'19)  & 34.65/0.9288   &30.55/0.8465   &29.25/0.8089  &28.74/0.8645  \\
		& SAN~\cite{dai2019second} (CVPR'19)   & \underline{34.75}/\underline{0.9300}  & 30.59/0.8476 & \underline{29.33}/\underline{0.8112} & 28.93/0.8671  \\
		%  DBPN~\cite{haris2018deep} & 6.0($\times$1.6) &34.7($\times$3.4) & 32.63 & 0.9051 \\
		& Pan~\cite{pan2020physics_sr} (AAAI'20)   &\underline{34.75}/0.9298 &30.61/0.8466 &29.29/0.8102 &28.97/0.8683   \\
		& HAN~\cite{niu2020single} (ECCV'20)     &\underline{34.75}/0.9299   &\textbf{30.67}/\underline{0.8483}   &29.32/0.8110  &\underline{29.10}/\underline{0.8705}  \\
		& \textbf{TSAN}  &34.64/0.9282   &30.52/0.8454  &29.20/0.8080 &28.55/0.8602  \\
		& \textbf{TSAN-L}  & \textbf{34.80/0.9301} & \underline{30.65}/\textbf{0.8486} & \textbf{29.34}/\textbf{0.8114} & \textbf{29.17}/\textbf{0.8720} \\
		
		\hline
		\multirow{10}{*}{$\times$4} & EDSR~\cite{lim2017enhanced} (CVPRW'17)  & 32.46/0.8968  & 28.80/0.7876 & 27.71/0.7420 & 26.64/0.8033  \\
		& RDN~\cite{zhang2018residual} (CVPR'18)  & 32.47/0.8990  & 28.81/0.7871 & 27.72/0.7417  & 26.61/0.8028 \\
		& DBPN~\cite{haris2018deep} (CVPR'18)  & 32.47/0.8980  & 28.82/0.7860 & 27.72/0.7400 & 26.38/0.7946 \\
		& RCAN~\cite{zhang2018image} (ECCV'18)  & 32.63/0.9002  & 28.87/\underline{0.7889}  & 27.77/0.7436 & 26.82/0.8087 \\		
		& RNAN~\cite{zhang2019residual} (ICLR'19)   & 32.49/0.8982 & 28.83/0.7878 & 27.72/0.7421 &26.61/0.8023  \\ 
		& SAN~\cite{dai2019second} (CVPR'19)  & \underline{32.64}/\underline{0.9003}  & \textbf{28.92}/0.7888 & 27.78/0.7436 & 26.79/0.8086  \\
		& Pan~\cite{pan2020physics_sr} (AAAI'20)    &32.56/0.8995 &28.80/0.7882 &27.73/0.7422 &26.72/0.8053  \\  
		& HAN~\cite{niu2020single} (ECCV'20)     &\underline{32.64}/0.9002   &28.90/\textbf{0.7890}  &\underline{27.80}/\underline{0.7442}  &\underline{26.85}/\underline{0.8094} \\
		& \textbf{TSAN}  & 32.40/0.8975  &28.73/0.7847   &27.67/0.7398  &26.39/0.7955  \\
		& \textbf{TSAN-L} & \textbf{32.65}/\textbf{0.9004} & \underline{28.91}/0.7888 & \textbf{27.81}/\textbf{0.7443} & \textbf{26.95}/\textbf{0.8110} \\
		
		\hline
		\hline
	\end{tabular}
	%  \vspace{-0.5cm}
\end{table*}

\subsection{\zjq{Comparisons with State-of-the-art Methods}}
%We performed several experiments and analysis to confirm the ability of our model.
We compare our proposed TSAN with eight state-of-the-art light-weighted methods (with $< 6$M parameters):  LapSRN~\cite{lai2017deep}, CARN~\cite{ahn2018fast}, SRMDNF~\cite{zhang2018learning},  NLRN~\cite{liu2018non}, MSRN~\cite{li2018multi}, FRSR~\cite{Soh_2019_CVPR}, OISR-RK2~\cite{he2019ode}, and LattienNet~\cite{luolatticenet}. %, and SRFBN~\cite{li2019srfbn}
%Table~\ref{tab:x234} shows quantitative comparisons for $\times$2, $\times$3, and $\times$4.
We summarize the quantitative comparisons for $\times$2, $\times$3, and $\times$4 in Table~\ref{tab:par}. As we can see, our TSAN achieves excellent performance on different datasets and different upsampling scales. We also visualize several examples with different upsampling scales in Figures~\ref{fig:visual_x2}, \ref{fig:visual_x3}, and \ref{fig:visual_x4}. % provide visual comparisons on different upsampling scales and different testsets.
Obviously, the SR images generated by other methods exhibit visible artifacts, while our proposed TSAN is able to generate a more visually pleasant image with clean details and sharp edges.
%For example, in Figure~\ref{fig:visual_x3}, the direction of the lines predicted by other methods is wrong and is ambiguous, and only our proposed model reconstructs clear textures and right direction.
This can be explained by the fact that rich attentive contextual features with long-range dependencies extracted by multiple MCABs in the LR space ensure a good initial coarse SR result, and then the SR result can be further improved by RAB.

%Note that we do not compare with six current state-of-the-art methods, {\em i.e.}, EDSR~\cite{lim2017enhanced}, RDN~\cite{zhang2018residual}, RCAN~\cite{zhang2018image}, SAN~\cite{dai2019second}, DBPN~\cite{haris2018deep}, and RNAN~\cite{zhang2019rnan} in Table~\ref{tab:x234}, because they have very large parameters. As we can see in Table~\ref{tab:params}, RDN, RCAN, SAN perform well due to their heavy models with large parameters. Our proposed TSAN still achieves comparable performance more efficiently with a much lighter network.
To further prove the effectiveness of the proposed model, we increase the number of MCAB 
to 13 (denoted as TSAN-L) to fairly compare with some methods with large parameters or heavy computations. We compare TSAN and TSAN-L with eight current state-of-the-art heavy-weight methods in Table~\ref{tab:params},  {\em i.e.}, EDSR~\cite{lim2017enhanced}, RDN~\cite{zhang2018residual}, DBPN~\cite{haris2018deep}, RCAN~\cite{zhang2018image}, RNAN~\cite{zhang2019residual}, SAN~\cite{dai2019second}, Pan~\cite{pan2020physics_sr}, and HAN~\cite{niu2020single}. 
We can see that our proposed TSAN still achieves comparable performance, and TSAN-L performs favorably against the state-of-the-art methods.  
For example, the proposed TSAN gains 0.11dB higher than EDSR~\cite{lim2017enhanced} on Set5~\cite{bevilacqua2012low} for $\times2$ scale, and the TSAN-L gains 0.13dB, 0.16dB, and 0.10dB higher than state-of-the-art methods RCAN~\cite{zhang2018image}, SAN~\cite{dai2019second}, and Pan~\cite{pan2020physics_sr} on Urban100~\cite{huang2015single} for $\times4$ scale, respectively. This observation suggests that our TSAN with cross-dimension interaction can make better use of more informative contextual features to boost reconstruction performance.
We also visualize several examples with different upsampling scales in Figures~\ref{fig:visual_heavyx2}, \ref{fig:visual_heavyx3}, and \ref{fig:visual_heavyx4}. As shown, most compared SR methods cannot recover the grids of buildings accurately and suffer from unpleasant blurring artifacts. In contrast, our TSAN-L obtains clearer details and reconstructs sharper high-frequency textures. Again, this strongly demonstrates the superiority of our method.
\def\wdenoising{0.5\linewidth}
\def\hdenoising{1.45in}
\begin{figure}[t]
	\setlength{\tabcolsep}{0.1pt}
	\centering
	%	\vspace{-0.2cm}
	\begin{tabular}{cc}

		\includegraphics[width=\wdenoising, height=\hdenoising]{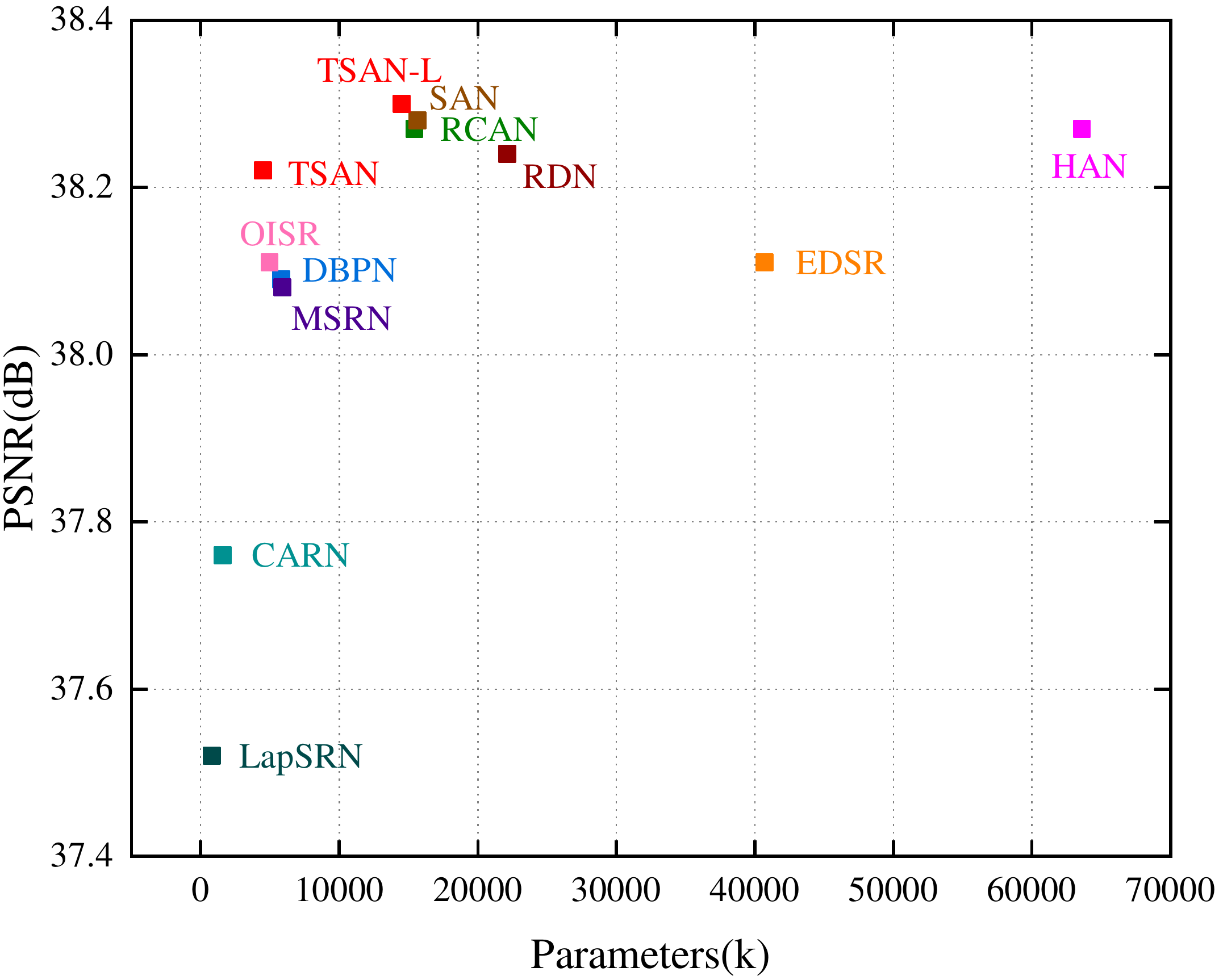}&
		\includegraphics[width=\wdenoising, height=\hdenoising]{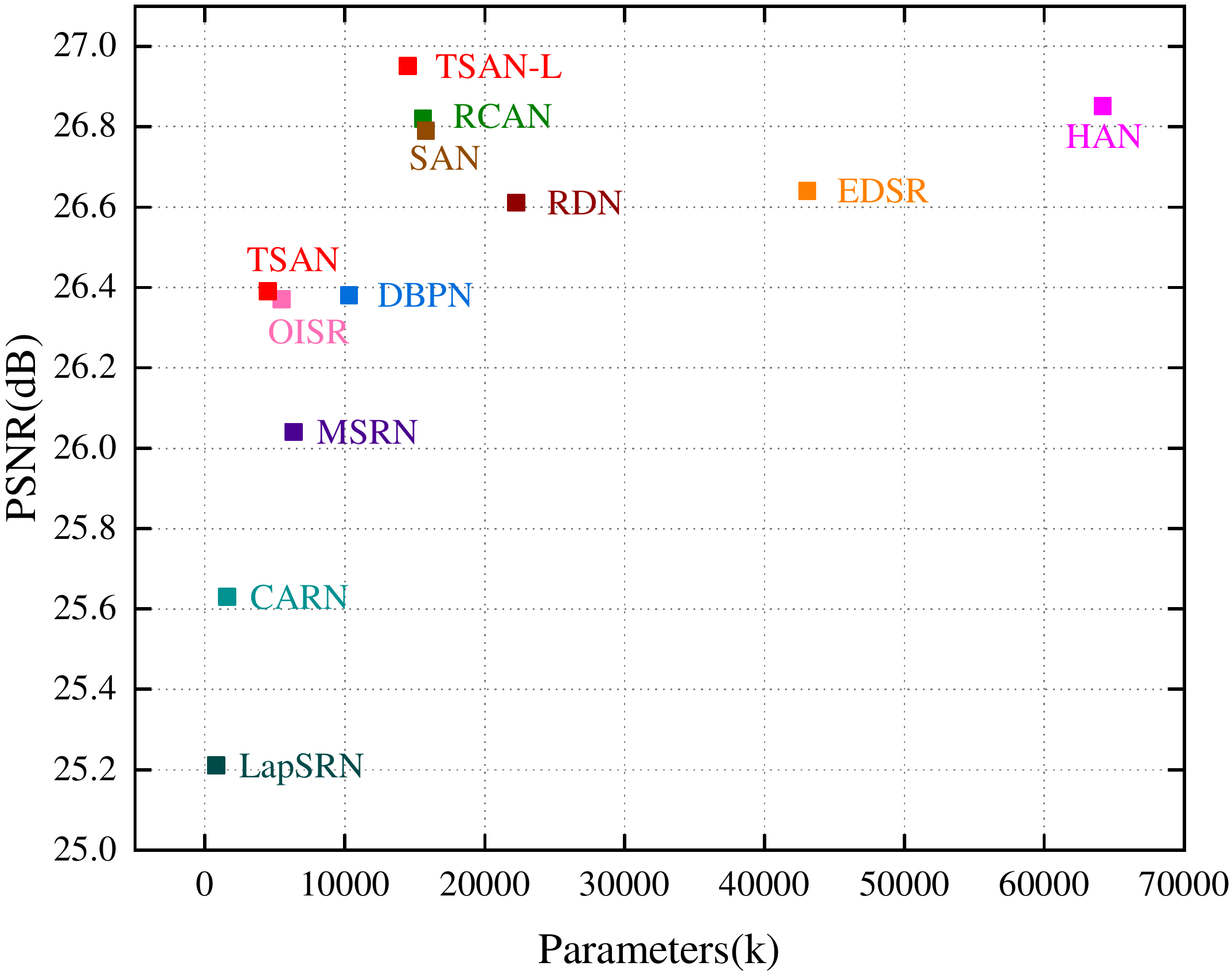}\\		
		(a)   & (b)  \\		
	\end{tabular}
	%	\vspace{-0.2cm}
	\caption{\zjq{PSNR performance versus number of parameters. (a) The results are evaluated on Set5 for a scale factor of $\times$2. (b) The results are evaluated on Urban100 for a scale factor of $\times$4. Our TSAN and TSAN-L has a better tradeoff between performance and model size.}}
	\label{fig:psnr_param}
%		\vspace{-0.3cm}
\end{figure}

We also compare the tradeoff between the performance and the number of network parameters from our TSAN network and existing methods. \zjq{Figure~\ref{fig:psnr_param} shows the PSNR performances
of 12 models versus the number of parameters, where the results are evaluated with Set5~\cite{bevilacqua2012low} and Urban100~\cite{huang2015single} datasets for 2$\times$ and 4$\times$ upscaling factors, respectively.} We can find that our TSAN and TSAN-L network significantly outperforms the relatively small models across all datasets and scales. 
Moreover, our TSAN-L network performs better than EDSR~\cite{lim2017enhanced} and RDN~\cite{zhang2018residual} across two scales but with about 65\% and 34\% fewer parameters on average, respectively. Furthermore, compared with RCAN~\cite{zhang2018image} and SAN~\cite{dai2019second}  on two upscaling factors, our TSAN-L has fewer parameters and achieves higher PSNR. 
\zjq{We further compute the FLOPs and provide the speed by assuming that the size of LR image is $48\times48$ and the scale factor is 2. For a fair comparison, all methods are tested on the same CPU. From Table~\ref{tab:SPEED}, we can see that our network has fewer FLOPs and faster inference speed than compared approaches.
These comparisons indicate that our proposed network has a better tradeoff between performance and model size.}

\setlength{\tabcolsep}{1.5pt}
\begin{table}[tbp]
	\centering
	%	\small
	\caption{\zjq{THE FLOPS AND INFERENCE TIME COMPARISONS OF OUR METHOD WITH FIVE STATE-OF-THE-ART NETWORKS.}}
	\label{tab:SPEED}
	\begin{tabular}{lcccccc}
		\hline
		\hline
		
		& EDSR~\cite{lim2017enhanced}  & MSRN~\cite{li2018multi} &RCAN~\cite{zhang2018image} & SAN~\cite{dai2019second} & HAN~\cite{niu2020single} & TSAN \\
		\hline
		FLOPs (G)  & 115.78 &13.67     &36.67 &30.04 &150.99 &10.11    \\
		Time (s)  & 17.00   &2.57  & 9.56    &10.26 &26.82 &2.34  \\
		\hline
		\hline
	\end{tabular}
	%  \vspace{-0.5cm}
\end{table}

% light x2
\def\wvisualimagex3{0.3\linewidth}
\def\hvisualimagex3{1.785in}
\def\wvisualx3{0.13\linewidth}
\def\hvisualx3{0.7in}
\begin{figure*}[ht!]
	\setlength{\tabcolsep}{2.3pt}
	\centering
	\begin{tabular}{cccccc}
		\multirow{4}{*}[42pt]{\includegraphics[width=\wvisualimagex3, height=\hvisualimagex3]{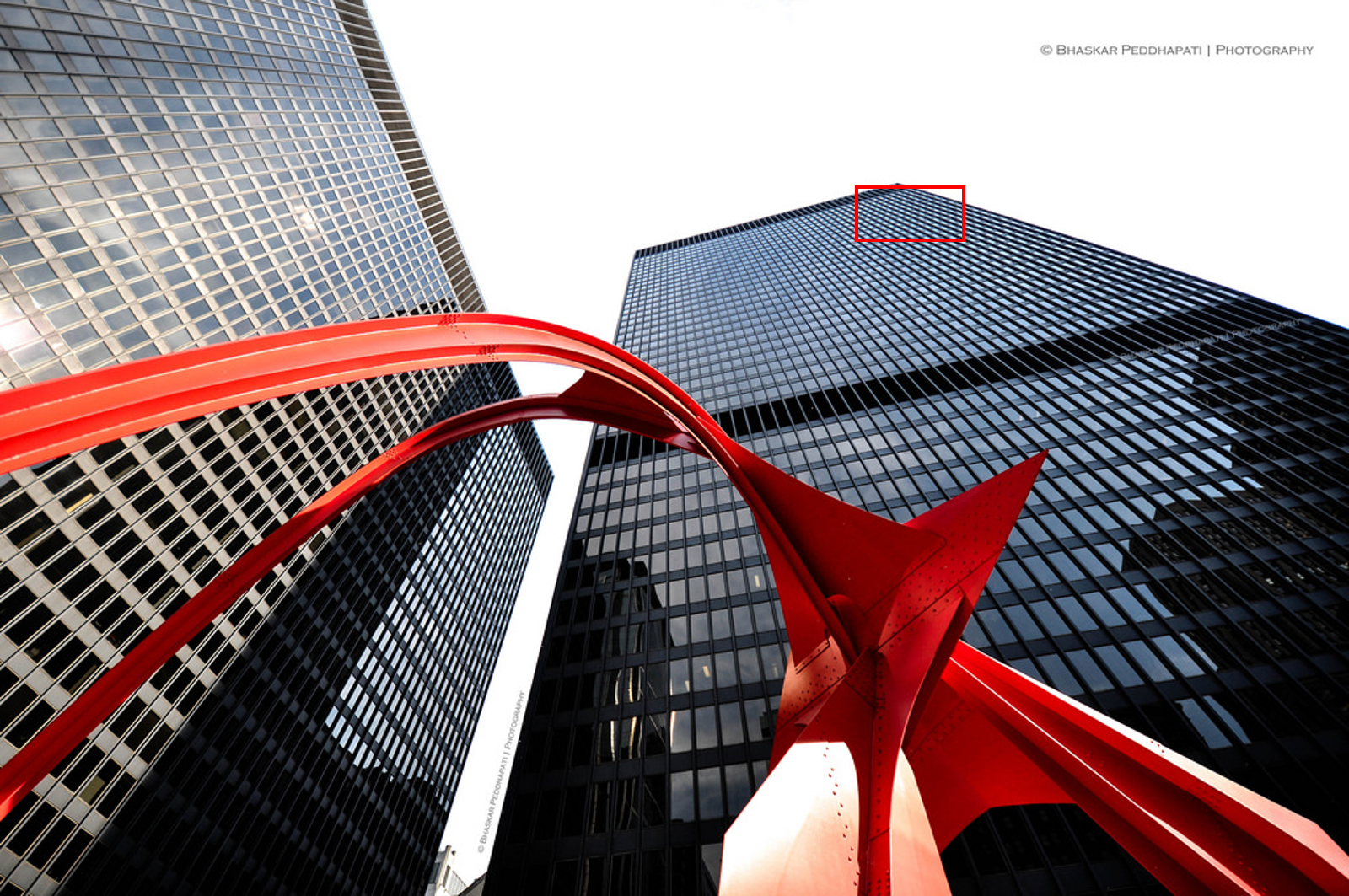}} &
		
		\includegraphics[width=\wvisualx3, height=\hvisualx3]{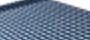}&
		\includegraphics[width=\wvisualx3, height=\hvisualx3]{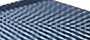}&
		\includegraphics[width=\wvisualx3, height=\hvisualx3]{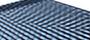}&
		\includegraphics[width=\wvisualx3, height=\hvisualx3]{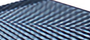}&
		\includegraphics[width=\wvisualx3, height=\hvisualx3]{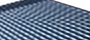}\\
		
		& \small{BIC} &  \small{LapSRN~\cite{lai2017deep}} &  \small{CARN~\cite{ahn2018fast}} &  \small{SRMDNF~\cite{zhang2018learning}} &  \small{NLRN~\cite{liu2018non}} \\
		& \small{(23.45/0.8528)} & \small{(26.69/0.9396)} & \small{(29.19/0.9638)} &\small{(27.95/0.9550)} & \small{(29.15/0.9622)} \\
		
		& \includegraphics[width=\wvisualx3, height=\hvisualx3]{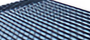}&
		\includegraphics[width=\wvisualx3, height=\hvisualx3]{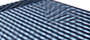}&
		\includegraphics[width=\wvisualx3, height=\hvisualx3]{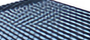}&
		\includegraphics[width=\wvisualx3, height=\hvisualx3]{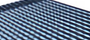}&
		\includegraphics[width=\wvisualx3, height=\hvisualx3]{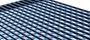}\\
		
		\multirow{2}{*}{Ground Truth} &  \small{MSRN~\cite{li2018multi}} &  \small{FRSR~\cite{Soh_2019_CVPR}} &  \small{OISR~\cite{he2019ode}} &  \small{TSAN} &  \small{GT} \\
		& \small{(29.84/0.9681)} & \small{(30.14/0.9689)} & \small{(29.96/0.9695)} &\small{\bf{(30.81/0.9729)}} & \small{(PSNR/SSIM)} \\
		
		\specialrule{0em}{1.5pt}{1.5pt}
		\multirow{4}{*}[42pt]{\includegraphics[width=\wvisualimagex3, height=\hvisualimagex3]{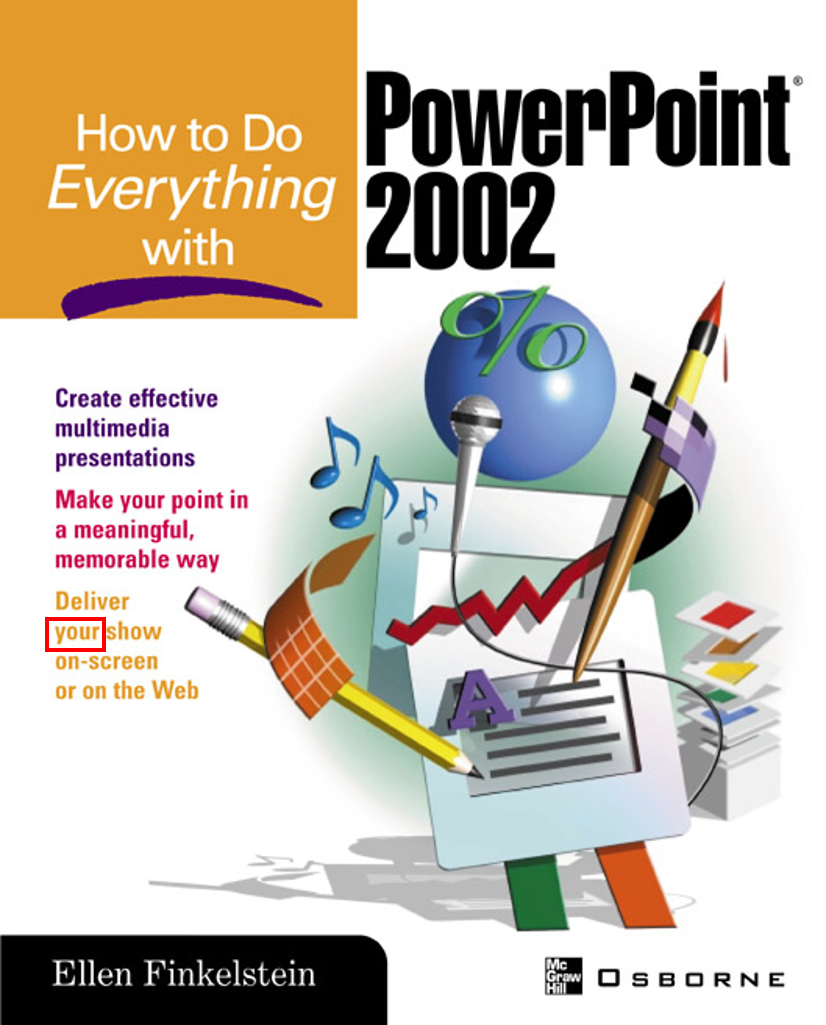}} &
		
		\includegraphics[width=\wvisualx3, height=\hvisualx3]{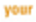}&
		\includegraphics[width=\wvisualx3, height=\hvisualx3]{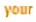}&
		\includegraphics[width=\wvisualx3, height=\hvisualx3]{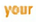}&
		\includegraphics[width=\wvisualx3, height=\hvisualx3]{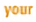}&
		\includegraphics[width=\wvisualx3, height=\hvisualx3]{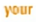}\\
		
		& \small{BIC} &  \small{LapSRN~\cite{lai2017deep}} &  \small{CARN~\cite{ahn2018fast}} &  \small{SRMDNF~\cite{zhang2018learning}} &  \small{NLRN~\cite{liu2018non}} \\
		& \small{(26.81/0.9466)} & \small{(32.82/0.9883)} & \small{(34.46/0.9912)} &\small{(34.23/0.9910)} & \small{(34.67/0.9919)} \\
		
		& \includegraphics[width=\wvisualx3, height=\hvisualx3]{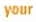}&
		\includegraphics[width=\wvisualx3, height=\hvisualx3]{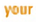}&
		\includegraphics[width=\wvisualx3, height=\hvisualx3]{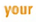}&
		\includegraphics[width=\wvisualx3, height=\hvisualx3]{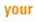}&
		\includegraphics[width=\wvisualx3, height=\hvisualx3]{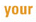}\\
		
		\multirow{2}{*}{Ground Truth} &  \small{MSRN~\cite{li2018multi}} &  \small{FRSR~\cite{Soh_2019_CVPR}} &  \small{OISR~\cite{he2019ode}} &  \small{TSAN} &  \small{GT} \\
		& \small{(35.12/0.9920)} & \small{(35.04/0.9920)} & \small{(35.21/0.9922)} &\small{\bf{(35.77/0.9932)}} & \small{(PSNR/SSIM)} \\
		
	\end{tabular}
	
	\caption{Visual comparison between our \textbf{TSAN }and other light-weighted methods on \emph{img062} from Urban100~\cite{huang2015single} and \emph{ppt} from Set14~\cite{zeyde2010single} with scale  $\times$2.}
	\label{fig:visual_x2}
	%   \vspace{-0.8cm}
\end{figure*}

%light x3
\def\wvisualimagex3{0.3\linewidth}
\def\hvisualimagex3{1.785in}
\def\wvisualx3{0.13\linewidth}
\def\hvisualx3{0.7in}
\begin{figure*}[ht!]
	\setlength{\tabcolsep}{2.3pt}
	\centering
	\begin{tabular}{cccccc}
		\multirow{4}{*}[42pt]{\includegraphics[width=\wvisualimagex3, height=\hvisualimagex3]{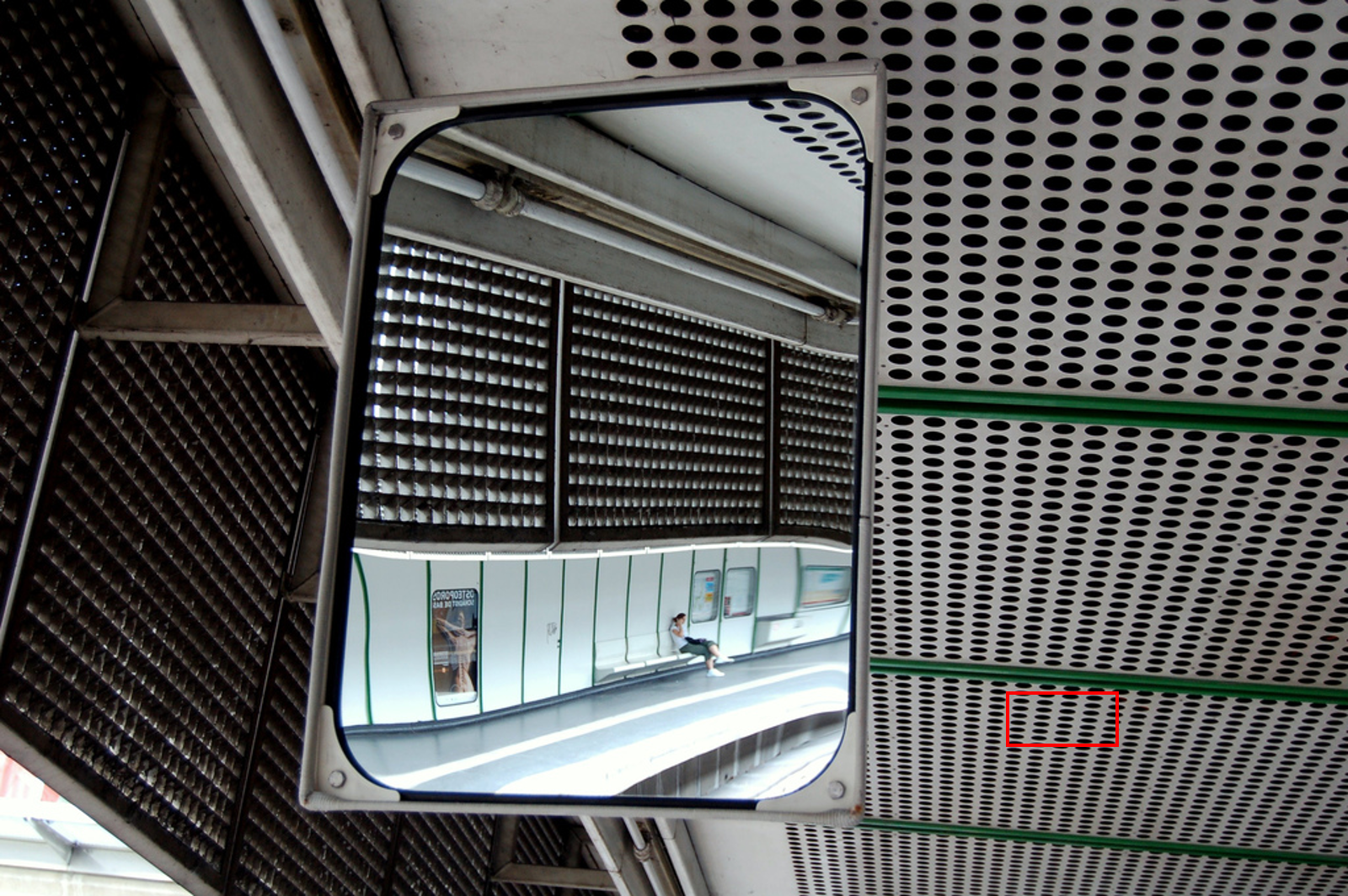}} &
		
		\includegraphics[width=\wvisualx3, height=\hvisualx3]{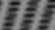}&
		\includegraphics[width=\wvisualx3, height=\hvisualx3]{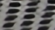}&
		\includegraphics[width=\wvisualx3, height=\hvisualx3]{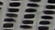}&
		\includegraphics[width=\wvisualx3, height=\hvisualx3]{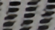}&
		\includegraphics[width=\wvisualx3, height=\hvisualx3]{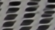}\\
		
		& \small{BIC} &  \small{LapSRN~\cite{lai2017deep}} &  \small{CARN~\cite{ahn2018fast}} &  \small{SRMDNF~\cite{zhang2018learning}} &  \small{NLRN~\cite{liu2018non}} \\
		& \small{(22.83/0.7840)} & \small{(24.80/0.8815)} &\small{(26.13/0.9083)} &\small{(25.33/0.8953)} &\small{(25.91/0.9055)} \\
		
		& \includegraphics[width=\wvisualx3, height=\hvisualx3]{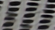}&
		\includegraphics[width=\wvisualx3, height=\hvisualx3]{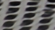}&
		\includegraphics[width=\wvisualx3, height=\hvisualx3]{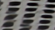}&
		\includegraphics[width=\wvisualx3, height=\hvisualx3]{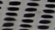}&
		\includegraphics[width=\wvisualx3, height=\hvisualx3]{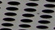}\\
		
		\multirow{2}{*}{Ground Truth} &  \small{MSRN~\cite{li2018multi}} &  \small{FRSR~\cite{Soh_2019_CVPR}} &  \small{OISR~\cite{he2019ode}} &  \small{TSAN} &  \small{GT} \\
		& \small{(25.97/0.9086)} & \small{(25.71/0.9051)} & \small{(25.97/0.9090)} &\small{\bf{(26.55/0.9161)}} & \small{(PSNR/SSIM)} \\
		
		\multirow{4}{*}[42pt]{\includegraphics[width=\wvisualimagex3, height=\hvisualimagex3]{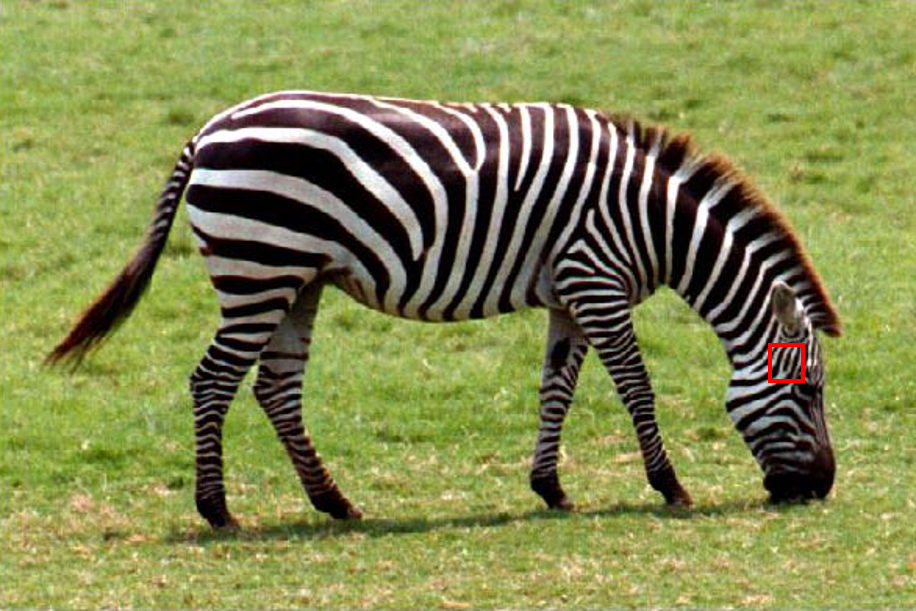}} &
		
		\includegraphics[width=\wvisualx3, height=\hvisualx3]{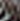}&
		\includegraphics[width=\wvisualx3, height=\hvisualx3]{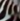}&
		\includegraphics[width=\wvisualx3, height=\hvisualx3]{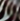}&
		\includegraphics[width=\wvisualx3, height=\hvisualx3]{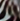}&
		\includegraphics[width=\wvisualx3, height=\hvisualx3]{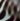}\\
		
		& \small{BIC} &  \small{LapSRN~\cite{lai2017deep}} &  \small{CARN~\cite{ahn2018fast}} &  \small{SRMDNF~\cite{zhang2018learning}} &  \small{NLRN~\cite{liu2018non}} \\
		& \small{(26.63/0.7953)} & \small{(29.87/0.8595)} & \small{(30.24/0.8657)} &\small{(30.01/0.8641)} & \small{(30.40/0.8665)} \\
		
		& \includegraphics[width=\wvisualx3, height=\hvisualx3]{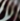}&
		\includegraphics[width=\wvisualx3, height=\hvisualx3]{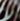}&
		\includegraphics[width=\wvisualx3, height=\hvisualx3]{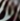}&
		\includegraphics[width=\wvisualx3, height=\hvisualx3]{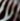}&
		\includegraphics[width=\wvisualx3, height=\hvisualx3]{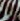}\\
		
		\multirow{2}{*}{Ground Truth} &  \small{MSRN~\cite{li2018multi}} &  \small{FRSR~\cite{Soh_2019_CVPR}} &  \small{OISR~\cite{he2019ode}} &  \small{TSAN} &  \small{GT} \\
		& \small{(30.40/0.8666)} & \small{(30.31/0.8660)} & \small{(30.42/0.8671)} &\small{\bf{(30.56/0.8684)}} & \small{(PSNR/SSIM)} \\

	\end{tabular}
	
	\caption{Visual comparison between our \textbf{TSAN} and other light-weighted methods on \emph{img004} from Urban100~\cite{huang2015single} and \emph{zebra} from Set14~\cite{zeyde2010single} with scale $\times$3.}
	\label{fig:visual_x3}
\end{figure*}

\def\wvisualimagex3{0.3\linewidth}
\def\hvisualimagex3{1.785in}
\def\wvisualx3{0.13\linewidth}
\def\hvisualx3{0.7in}
\begin{figure*}[ht!]
	\setlength{\tabcolsep}{2.3pt}
	\centering
	\begin{tabular}{cccccc}
		\multirow{4}{*}[42pt]{\includegraphics[width=\wvisualimagex3, height=\hvisualimagex3]{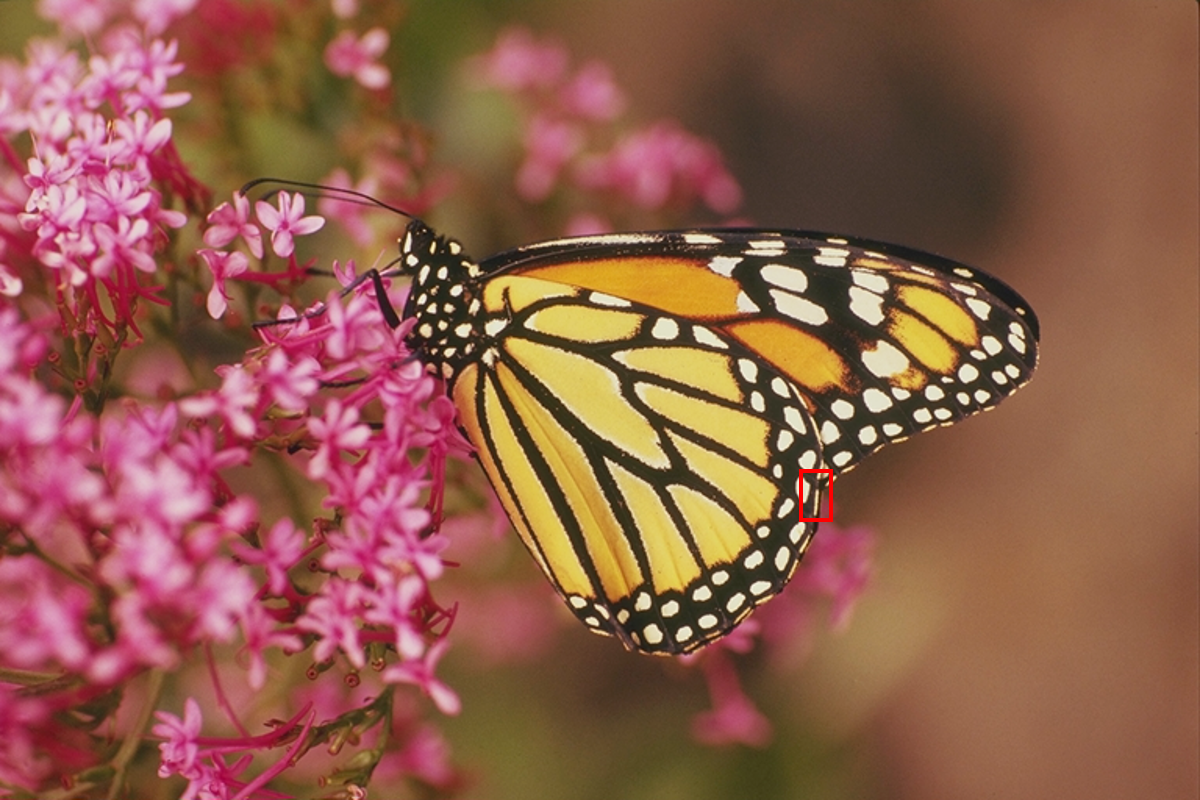}} &
		
		\includegraphics[width=\wvisualx3, height=\hvisualx3]{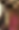}&
		\includegraphics[width=\wvisualx3, height=\hvisualx3]{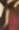}&
		\includegraphics[width=\wvisualx3, height=\hvisualx3]{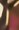}&
		\includegraphics[width=\wvisualx3, height=\hvisualx3]{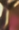}&
		\includegraphics[width=\wvisualx3, height=\hvisualx3]{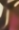}\\
		
		& \small{BIC} &  \small{LapSRN~\cite{lai2017deep}} &  \small{CARN~\cite{ahn2018fast}} &  \small{SRMDNF~\cite{zhang2018learning}} &  \small{NLRN~\cite{liu2018non}} \\
		&  \small{(27.46/0.8814)} &  \small{(31.63/0.9357)} &  \small{(33.03/0.9447)} & \small{(32.36/0.9405)} &  \small{(32.46/0.9405)} \\
		
		& \includegraphics[width=\wvisualx3, height=\hvisualx3]{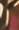}&
		\includegraphics[width=\wvisualx3, height=\hvisualx3]{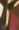}&
		\includegraphics[width=\wvisualx3, height=\hvisualx3]{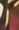}&
		\includegraphics[width=\wvisualx3, height=\hvisualx3]{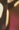}&
		\includegraphics[width=\wvisualx3, height=\hvisualx3]{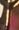}\\
		
		\multirow{2}{*}{Ground Truth} &  \small{MSRN~\cite{li2018multi}} &  \small{FRSR~\cite{Soh_2019_CVPR}} &  \small{OISR~\cite{he2019ode}} &  \small{TSAN} &  \small{GT} \\
		&  \small{(33.15/0.9459)} &  \small{(33.15/0.9455)} &  \small{(33.37/0.9471)} & \small{\bf{(33.48/0.9480)}} &  \small{(PSNR/SSIM)} \\
		
		\multirow{4}{*}[42pt]{\includegraphics[width=\wvisualimagex3, height=\hvisualimagex3]{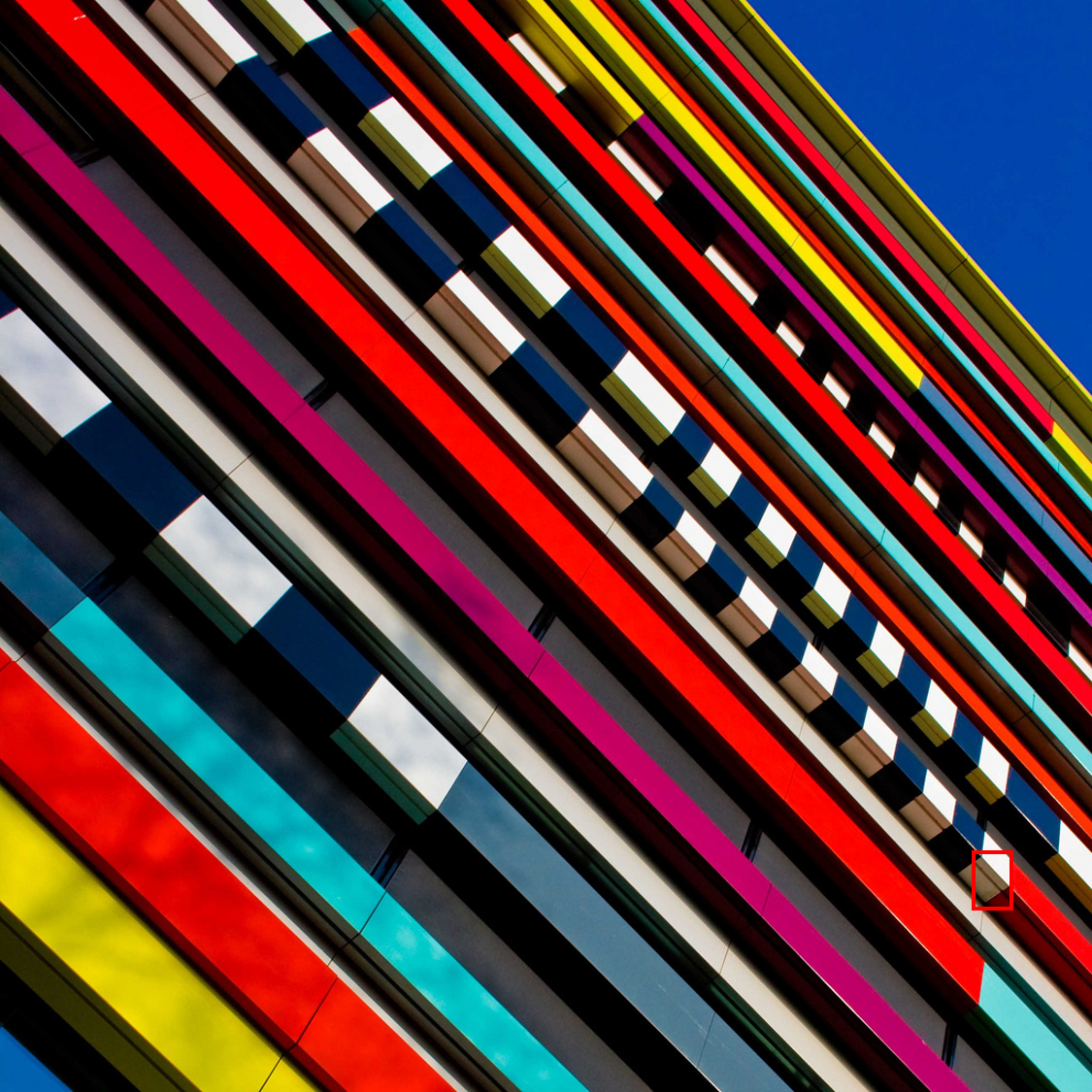}} &
		
		\includegraphics[width=\wvisualx3, height=\hvisualx3]{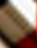}&
		\includegraphics[width=\wvisualx3, height=\hvisualx3]{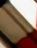}&
		\includegraphics[width=\wvisualx3, height=\hvisualx3]{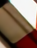}&
		\includegraphics[width=\wvisualx3, height=\hvisualx3]{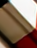}&
		\includegraphics[width=\wvisualx3, height=\hvisualx3]{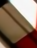}\\
		
		&  \small{BIC} &  \small{LapSRN~\cite{lai2017deep}} &  \small{CARN~\cite{ahn2018fast}} &  \small{SRMDNF~\cite{zhang2018learning}} &  \small{NLRN~\cite{liu2018non}} \\
		&  \small{(28.57/0.8930)} &  \small{(36.37/0.9685)} &  \small{(38.12/0.9752)} & \small{(36.38/0.9681)} &  \small{(37.46/0.9719)} \\
		
		& \includegraphics[width=\wvisualx3, height=\hvisualx3]{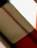}&
		\includegraphics[width=\wvisualx3, height=\hvisualx3]{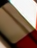}&
		\includegraphics[width=\wvisualx3, height=\hvisualx3]{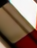}&
		\includegraphics[width=\wvisualx3, height=\hvisualx3]{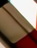}&
		\includegraphics[width=\wvisualx3, height=\hvisualx3]{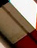}\\
		
		\multirow{2}{*}{Ground Truth} &  \small{MSRN~\cite{li2018multi}} &  \small{FRSR~\cite{Soh_2019_CVPR}} &  \small{OISR~\cite{he2019ode}} &  \small{TSAN} &  \small{GT} \\
		&  \small{(38.17/0.9750)} &  \small{(38.16/0.9738)} &  \small{(38.24/0.9755)} & \small{\bf{(38.62/0.9773)}} &  \small{(PSNR/SSIM)} \\
		
	\end{tabular}
	
	\caption{Visual comparison between our \textbf{TSAN} and other light-weighted methods  on \emph{monarch} from Set14~\cite{zeyde2010single} and \emph{img081} from Urban100~\cite{huang2015single} with scale $\times$4.}
	\label{fig:visual_x4}
\end{figure*}

%heavy x2
\def\wvisualimagex3{0.3\linewidth}
\def\hvisualimagex3{1.785in}
\def\wvisualx3{0.16\linewidth}
\def\hvisualx3{0.7in}
\begin{figure*}[ht!]
	\setlength{\tabcolsep}{2.3pt}
	\centering
	\begin{tabular}{ccccc}
		\multirow{4}{*}[42pt]{\includegraphics[width=\wvisualimagex3, height=\hvisualimagex3]{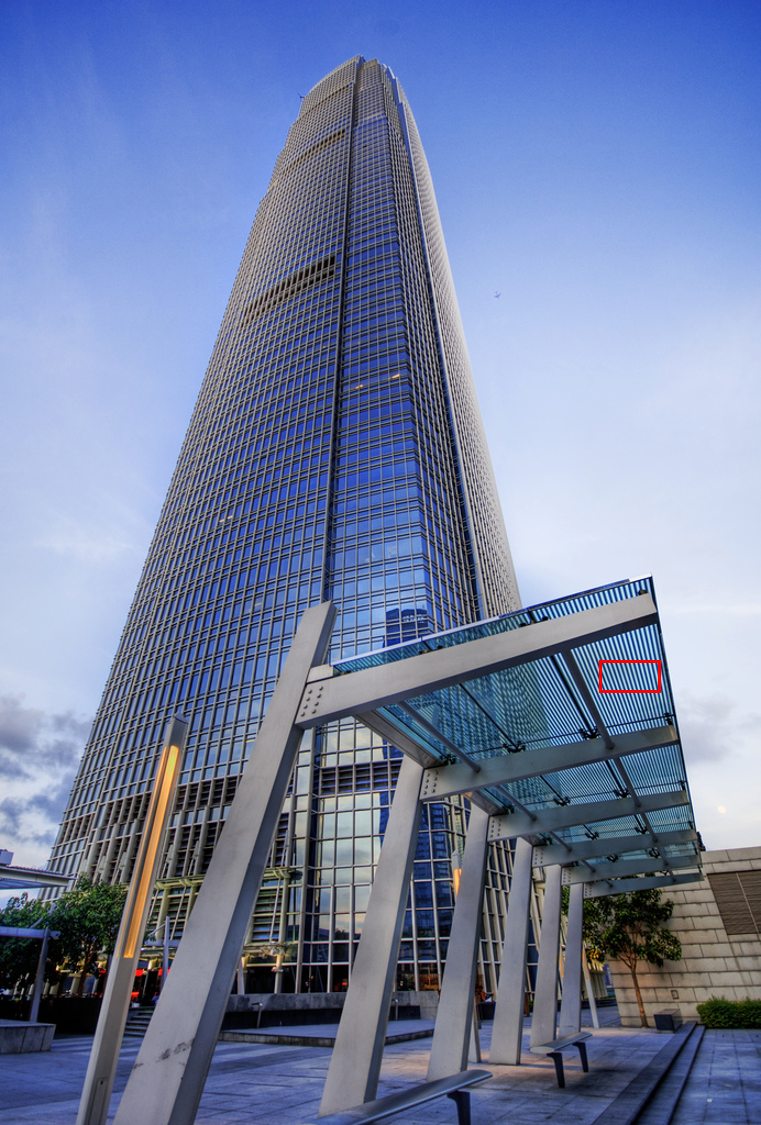}} &
		
		\includegraphics[width=\wvisualx3, height=\hvisualx3]{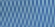}&
		\includegraphics[width=\wvisualx3, height=\hvisualx3]{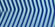}&
		\includegraphics[width=\wvisualx3, height=\hvisualx3]{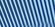}&
		\includegraphics[width=\wvisualx3, height=\hvisualx3]{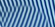}\\
		
		& \small{BIC} & \small{RDN~\cite{zhang2018residual}} & \small{RCAN~\cite{zhang2018image}} & \small{RNAN~\cite{zhang2019rnan}} \\
		&\small{(25.24/0.8343)}  & \small{(28.99/0.9417)} &\small{(30.59/0.9515)} & \small{(29.56/0.9436)} \\
		
		& \includegraphics[width=\wvisualx3, height=\hvisualx3]{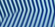}&
		\includegraphics[width=\wvisualx3, height=\hvisualx3]{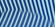}&
		\includegraphics[width=\wvisualx3, height=\hvisualx3]{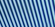}&
		\includegraphics[width=\wvisualx3, height=\hvisualx3]{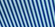}\\
		
		\multirow{2}{*}{Ground Truth} &  \small{SAN~\cite{dai2019second}} & \small{Pan~\cite{pan2020physics_sr}} & \small{TSAN} & \small{GT} \\
		& \small{(29.01/0.9414)} & \small{(29.51/0.9461)} &\small{\bf{(31.03/0.9540)}} & \small{(PSNR/SSIM)} \\
		
		\multirow{4}{*}[42pt]{\includegraphics[width=\wvisualimagex3, height=\hvisualimagex3]{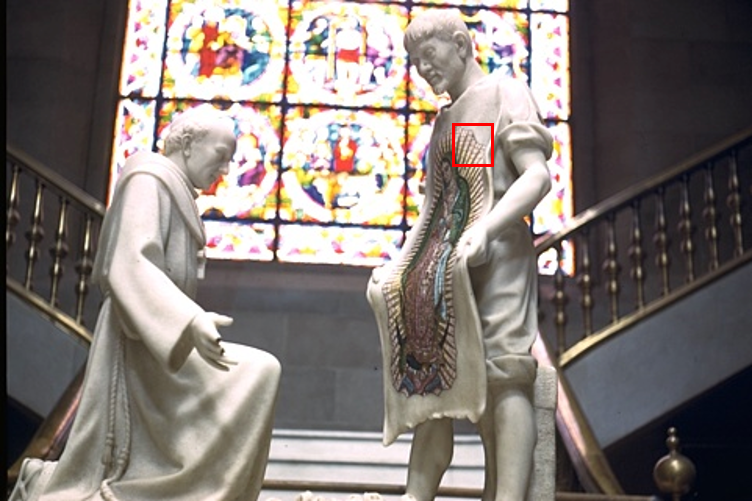}} &
		
		\includegraphics[width=\wvisualx3, height=\hvisualx3]{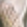}&
		\includegraphics[width=\wvisualx3, height=\hvisualx3]{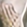}&
		\includegraphics[width=\wvisualx3, height=\hvisualx3]{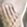}&
		\includegraphics[width=\wvisualx3, height=\hvisualx3]{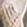}\\
		
		& \small{BIC} & \small{RDN~\cite{zhang2018residual}} & \small{RCAN~\cite{zhang2018image}} & \small{RNAN~\cite{zhang2019rnan}} \\
		&\small{(28.86/0.9267)}  & \small{(34.62/0.9733)}&\small{(34.77/0.9741)} & \small{(34.54/0.9731)} \\
		
		& \includegraphics[width=\wvisualx3, height=\hvisualx3]{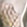}&
		\includegraphics[width=\wvisualx3, height=\hvisualx3]{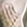}&
		\includegraphics[width=\wvisualx3, height=\hvisualx3]{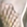}&
		\includegraphics[width=\wvisualx3, height=\hvisualx3]{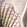}\\
		
		\multirow{2}{*}{Ground Truth} &  \small{SAN~\cite{dai2019second}} & \small{Pan~\cite{pan2020physics_sr}} & \small{TSAN-L} & \small{GT} \\
		& \small{(34.93/0.9744)} & \small{(34.79/0.9471)} &\small{\bf{(35.03/0.9748)}} & \small{(PSNR/SSIM)} \\

	\end{tabular}
	
	\caption{\zjq{Visual comparison between our \textbf{TSAN-L} and several heavy-weighted methods  on \emph{img046} from Urban100~\cite{huang2015single} and \emph{24077} from BSDS100~\cite{arbelaez2011contour} with scale $\times$2.}}
	\label{fig:visual_heavyx2}
\end{figure*}

% heavy x3
\def\wvisualimagex3{0.3\linewidth}
\def\hvisualimagex3{1.785in}
\def\wvisualx3{0.16\linewidth}
\def\hvisualx3{0.7in}
\begin{figure*}[ht!]
	\setlength{\tabcolsep}{2.3pt}
	\centering
	\begin{tabular}{ccccc}
		\multirow{4}{*}[42.0pt]{\includegraphics[width=\wvisualimagex3, height=\hvisualimagex3]{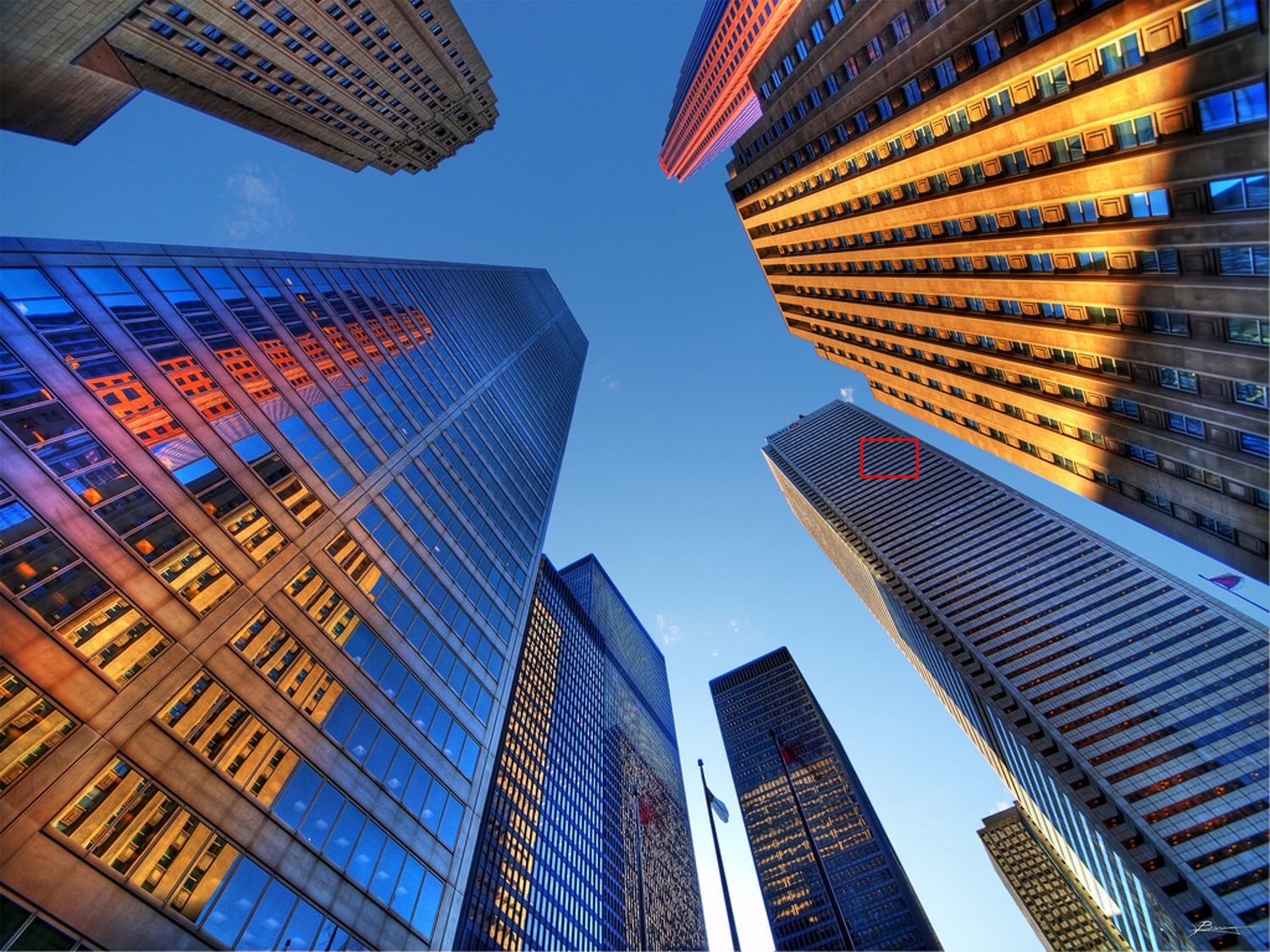}} &
		
		\includegraphics[width=\wvisualx3, height=\hvisualx3]{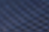}&
		\includegraphics[width=\wvisualx3, height=\hvisualx3]{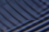}&
		\includegraphics[width=\wvisualx3, height=\hvisualx3]{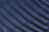}&
		\includegraphics[width=\wvisualx3, height=\hvisualx3]{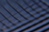}\\
		
		%		& \footnotesize{BIC} & \footnotesize{RDN~\cite{zhang2018residual}} & \footnotesize{RCAN~\cite{zhang2018image}} & \footnotesize{RNAN~\cite{zhang2019rnan}}  \\
		%		& \footnotesize{(23.45/0.6901)}  & \footnotesize{(25.95/0.8370)} &\footnotesize{(26.08/0.8393)} & \footnotesize{(25.81/0.8322)} \\
		& \small{BIC} & \small{RDN~\cite{zhang2018residual}} & \small{RCAN~\cite{zhang2018image}} & \small{RNAN~\cite{zhang2019rnan}} \\
		& \small{(23.45/0.6901)}  & \small{(25.95/0.8370)} &\small{(26.08/0.8393)} & \small{(25.81/0.8322)} \\
		
		& \includegraphics[width=\wvisualx3, height=\hvisualx3]{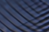}&
		\includegraphics[width=\wvisualx3, height=\hvisualx3]{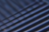}&
		\includegraphics[width=\wvisualx3, height=\hvisualx3]{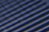}&
		\includegraphics[width=\wvisualx3, height=\hvisualx3]{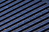}\\
		
		\multirow{2}{*}{Ground Truth} &  \small{SAN~\cite{dai2019second}} & \small{Pan~\cite{pan2020physics_sr}} & \small{TSAN} & \small{GT} \\
		& \small{(25.62/0.8348)} & \small{(26.01/0.8406)} &\small{\bf{(26.21/0.8447)}} & \small{(PSNR/SSIM)} \\
		
		\multirow{4}{*}[42.0pt]{\includegraphics[width=\wvisualimagex3, height=\hvisualimagex3]{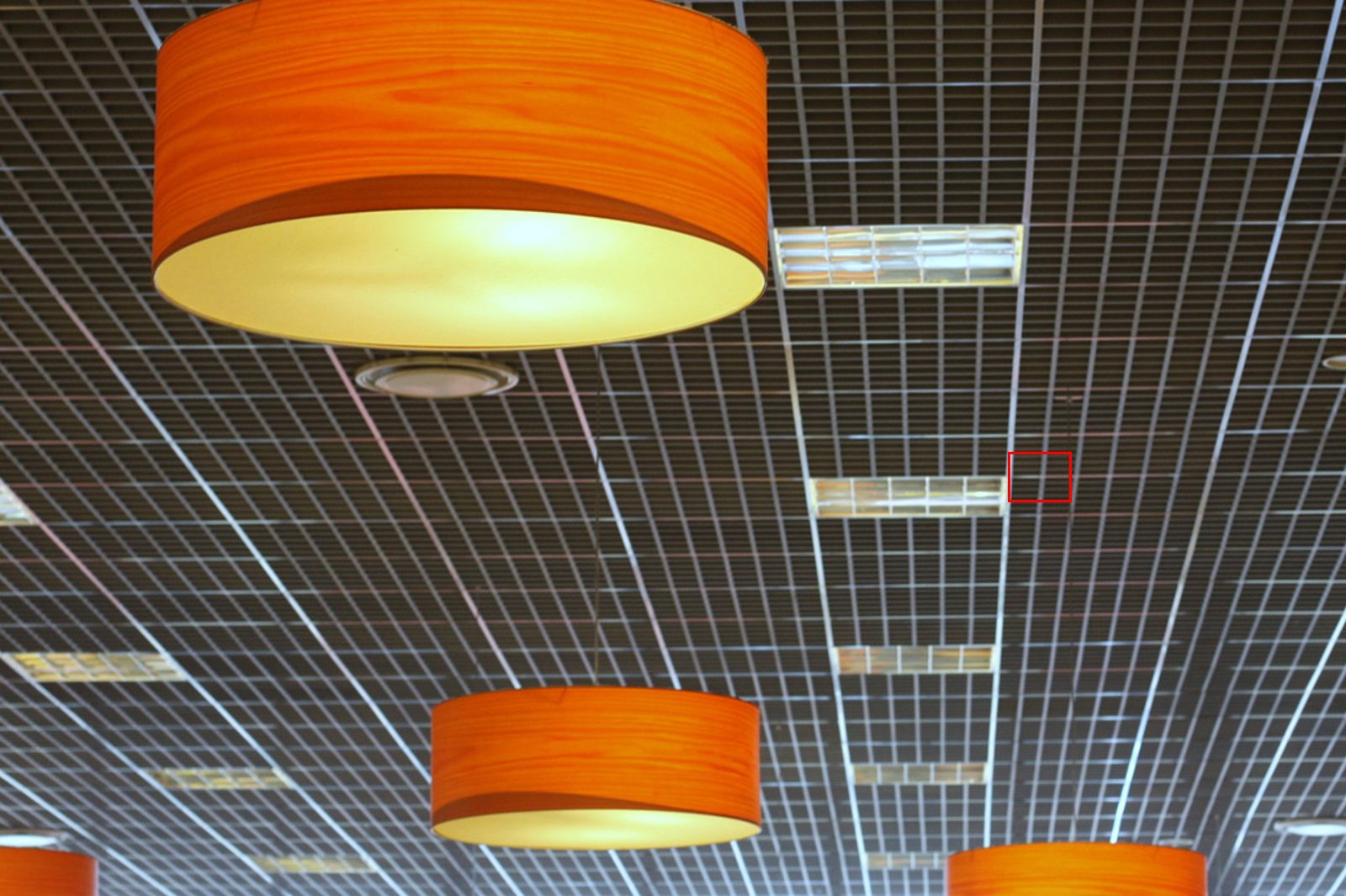}} &
		
		\includegraphics[width=\wvisualx3, height=\hvisualx3]{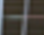}&
		\includegraphics[width=\wvisualx3, height=\hvisualx3]{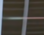}&
		\includegraphics[width=\wvisualx3, height=\hvisualx3]{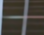}&
		\includegraphics[width=\wvisualx3, height=\hvisualx3]{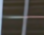}\\
		
		& \small{BIC} & \small{RDN~\cite{zhang2018residual}} & \small{RCAN~\cite{zhang2018image}} & \small{RNAN~\cite{zhang2019rnan}} \\
		&\small{(29.80/0.8388)}  &\small{(36.81/0.9538)} &\small{(36.89/0.9577)} & \small{(37.07/0.9534)} \\
		
		& \includegraphics[width=\wvisualx3, height=\hvisualx3]{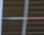}&
		\includegraphics[width=\wvisualx3, height=\hvisualx3]{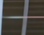}&
		\includegraphics[width=\wvisualx3, height=\hvisualx3]{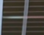}&
		\includegraphics[width=\wvisualx3, height=\hvisualx3]{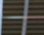}\\
		
		\multirow{2}{*}{Ground Truth} &  \small{SAN~\cite{dai2019second}} & \small{Pan~\cite{pan2020physics_sr}} & \small{TSAN-L} & \small{GT} \\
		&\small{(37.50/0.9595)} &\small{(35.43/0.9460)} &\small{\bf{(37.51/0.9601)}} & \small{(PSNR/SSIM)} \\
		
	\end{tabular}
	
	\caption{\zjq{Visual comparison between our \textbf{TSAN-L} and several heavy-weighted methods  on \emph{img012} from Urban100~\cite{huang2015single} and \emph{img044} from Urban100~\cite{huang2015single} with scale $\times$3.}}
	\label{fig:visual_heavyx3}
\end{figure*}

%heavy x4
\def\wvisualimagex3{0.3\linewidth}
\def\hvisualimagex3{1.785in}
\def\wvisualx3{0.16\linewidth}
\def\hvisualx3{0.7in}
\begin{figure*}[ht!]
	\setlength{\tabcolsep}{2.3pt}
	\centering
	\begin{tabular}{ccccc}
		\multirow{4}{*}[42pt]{\includegraphics[width=\wvisualimagex3, height=\hvisualimagex3]{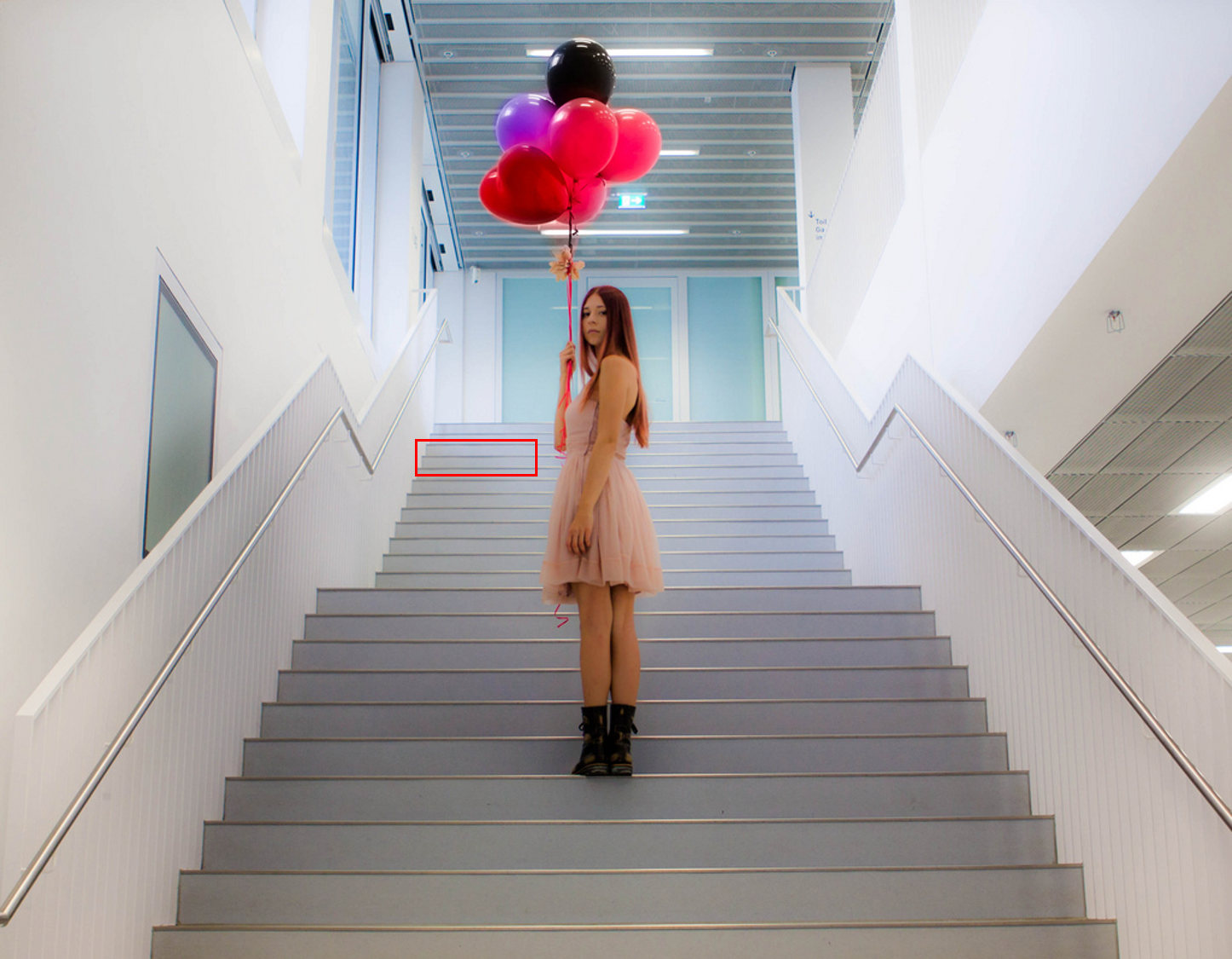}} &
		
		\includegraphics[width=\wvisualx3, height=\hvisualx3]{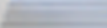}&
		\includegraphics[width=\wvisualx3, height=\hvisualx3]{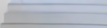}&
		\includegraphics[width=\wvisualx3, height=\hvisualx3]{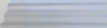}&
		\includegraphics[width=\wvisualx3, height=\hvisualx3]{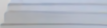}\\
		
		& \small{BIC} & \small{RDN~\cite{zhang2018residual}} & \small{RCAN~\cite{zhang2018image}} & \small{RNAN~\cite{zhang2019rnan}} \\
		& \small{(30.88/0.8560)}  & \small{(34.32/0.9306)} &\small{(34.59/0.9347)} & \small{(34.78/0.9351)} \\
		
		& \includegraphics[width=\wvisualx3, height=\hvisualx3]{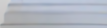}&
		\includegraphics[width=\wvisualx3, height=\hvisualx3]{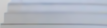}&
		\includegraphics[width=\wvisualx3, height=\hvisualx3]{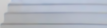}&
		\includegraphics[width=\wvisualx3, height=\hvisualx3]{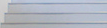}\\
		
		\multirow{2}{*}{Ground Truth} &  \small{SAN~\cite{dai2019second}} & \small{Pan~\cite{pan2020physics_sr}} & \small{TSAN} & \small{GT} \\
		& \small{(34.51/0.9284)} & \small{(34.63/0.9333)} &\small{\bf{(34.85/0.9382)}} & \small{(PSNR/SSIM)} \\
		
		\multirow{4}{*}[42pt]{\includegraphics[width=\wvisualimagex3, height=\hvisualimagex3]{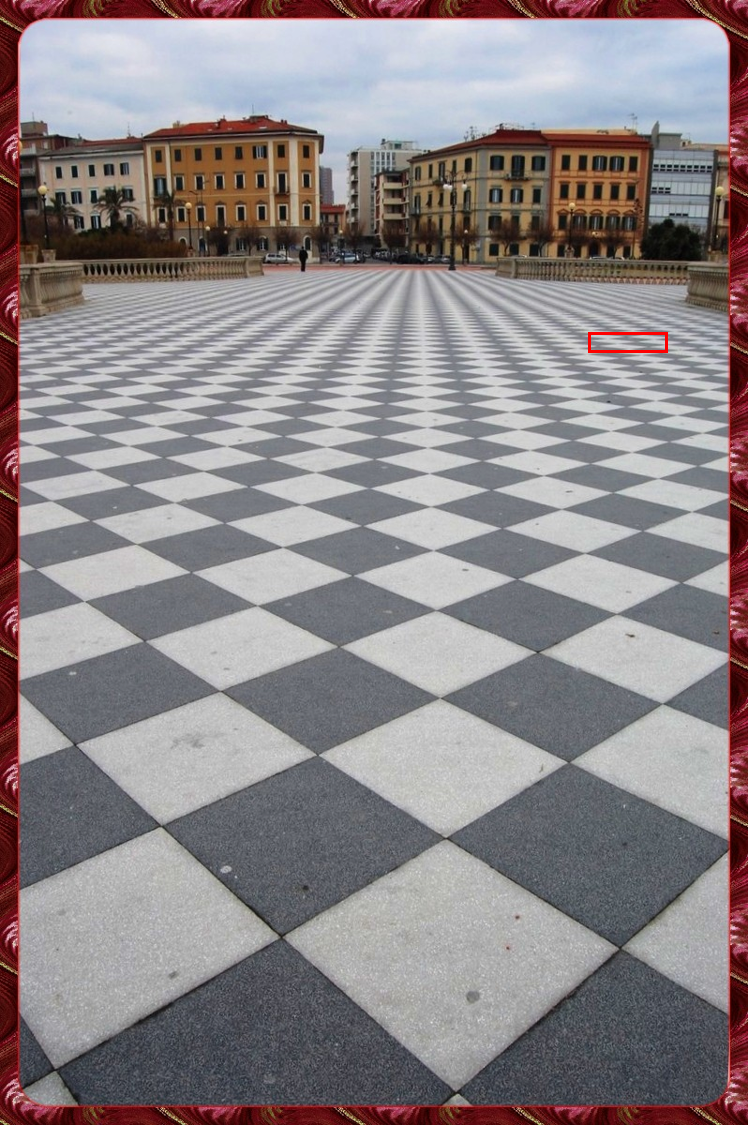}} &
		
		\includegraphics[width=\wvisualx3, height=\hvisualx3]{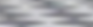}&
		\includegraphics[width=\wvisualx3, height=\hvisualx3]{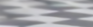}&
		\includegraphics[width=\wvisualx3, height=\hvisualx3]{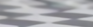}&
		\includegraphics[width=\wvisualx3, height=\hvisualx3]{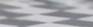}\\
		
		& \small{BIC} & \small{RDN~\cite{zhang2018residual}} & \small{RCAN~\cite{zhang2018image}} & \small{RNAN~\cite{zhang2019rnan}} \\
		& \small{(27.72/0.7199)} & \small{(30.15/0.8012)} &\small{(30.11/0.8011)} & \small{(30.16/0.8008)} \\
		
		& \includegraphics[width=\wvisualx3, height=\hvisualx3]{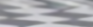}&
		\includegraphics[width=\wvisualx3, height=\hvisualx3]{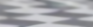}&
		\includegraphics[width=\wvisualx3, height=\hvisualx3]{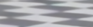}&
		\includegraphics[width=\wvisualx3, height=\hvisualx3]{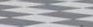}\\
		
		\multirow{2}{*}{Ground Truth} &  \small{SAN~\cite{dai2019second}} & \small{Pan~\cite{pan2020physics_sr}} & \small{TSAN-L} & \small{GT} \\
		& \small{(30.09/0.8004)} & \small{(30.08/0.7998)} &\small{\bf{(30.40/0.8054)}} & \small{(PSNR/SSIM)} \\
		
	\end{tabular}
	
	\caption{\zjq{Visual comparison between our \textbf{TSAN-L} and several heavy-weighted methods  on \emph{img009} from Urban100~\cite{huang2015single} and \emph{img021} from Urban100~\cite{huang2015single} with scale $\times$4.}}
	\label{fig:visual_heavyx4}
\end{figure*}

%% file: conclusion.tex
\section{conclusion}
In this paper, we propose a novel and light-weighted TSAN  for SISR in a coarse-to-fine fashion to utilize the attentive contextual information with cross-dimension interaction and emphasize the reconstruction process on both LR and SR space. The DRB with the well designed compact structure can increase the receptive field and get more contextual features. The MCAB  can  effectively  extract attentive contextual features by exploiting inter-dependencies between different dimensions to obtain the coarse result. 
Also, an RAB is proposed to focus on extracting essential HR space features after upsampling to refine the coarse result. Extensive evaluations on the benchmark datasets have demonstrated the efficacy of our proposed TSAN in terms of metric accuracy and visual effects.